\documentclass[a4paper,11pt]{article}
\pdfoutput=1 
\usepackage{jheppub} 
\makeatletter \renewcommand{\@fpheader}{} \makeatother

\UseRawInputEncoding
\usepackage{booktabs}
\usepackage{diagbox} 
\usepackage{amsmath,amsfonts,amssymb,graphics}
\usepackage{hyperref}
\usepackage[normalem]{ulem}
\usepackage{url}
\usepackage{graphicx}
\usepackage{cancel}
\usepackage{float}
 \usepackage{comment}
\usepackage{mathrsfs}
\usepackage{bbm}
\usepackage{pifont}
\usepackage{braket}
\usepackage{siunitx}
\usepackage{slashed}
\usepackage{mathtools}
\usepackage{bm}
\usepackage{xspace}
\usepackage{booktabs}
\usepackage{csquotes}
\usepackage[dvipsnames,svgnames]{xcolor}
\usepackage{tikz}
\usetikzlibrary{positioning,shapes,arrows}
\usetikzlibrary{decorations.pathmorphing}
\usetikzlibrary{decorations.markings}
\usepackage{standalone}
\usepackage{lmodern}
\usepackage{adjustbox}
\usepackage{makecell}

\newcommand{\SU}{{\rm SU}}
\newcommand{\nubar}{\bar{\nu}}

\title{Signatures of Neutrinophilic Pion Dark Matter} 

\author[a]{Claire Chevallier,}
\emailAdd{chevallier@ijclab.in2p3.fr}

\author[b,c]{Ji-Heng Guo,}
\emailAdd{guojiheng@buaa.edu.cn}

\author[b]{Giacomo Landini,}
\emailAdd{glandini@lnf.infn.it}

\affiliation[a]{Universit\'{e} Paris-Saclay, CNRS/IN2P3, IJCLab, 91405 Orsay, France}
\affiliation[b]{Istituto Nazionale di Fisica Nucleare, Laboratori Nazionali di Frascati, C.P. 13, 00044 Frascati, Italy}
\affiliation[c]{School of Physics, Beihang University, Beijing 102206, China}

\begin{document}

\abstract{

We study the phenomenology of pion dark matter coupled to neutrinos in a confining sector that dynamically generates
both small neutrino masses and dark matter mass from gauge confinement. Dark matter semi-annihilation to active neutrinos or annihilation to light sterile neutrinos can reproduce the observed relic abundance through thermal freeze-out while satisfying constraints from self-interactions. In the first scenario, semi-annihilations of sub-GeV dark matter in the galactic halo produce a flux of antineutrinos which can provide a viable interpretation for the mild electron antineutrino excess reported by the Super-Kamiokande collaboration. We assess the potential 
of the JUNO detector to test this interpretation. In the second scenario, light sterile neutrinos can leave signatures through displaced vertex searches at  accelerator facilities.

}

\maketitle

\section{Introduction}\label{sec:intro}

The existence of Dark Matter (DM) is supported by a wide range of astrophysical and cosmological observations, yet its microscopic nature and interactions with Standard Model particles remain unknown.  One intriguing possibility is that DM interacts predominantly with neutrinos, suppressing its signatures in searches involving charged particles. Such scenarios also offer the opportunity to connect DM physics to the generation of neutrino masses and to investigate the dark sector through its interactions with neutrinos: DM annihilations or decays can generate astrophysical neutrino fluxes, while searches for sterile neutrinos mediating the interactions to the Standard Model provide complementary tests at accelerator facilities.

The smallness of active neutrino masses is often explained by the introduction of a very heavy mediator. An interesting alternative is the inverse seesaw mechanism~\cite{Mohapatra:1986bd,Gonzalez-Garcia:1988okv}, in which mildly heavy vector-like fermion mediators, often at the TeV scale or below, are accompanied by a small explicit lepton-number violating term.
Recently, Ref.~\cite{Berbig:2024uwm} proposed a framework in which new confining dynamics generates the small lepton-number-violating parameter required by the inverse seesaw mechanism. This connects the origin of small neutrino masses to a dark sector containing composite DM candidates. While that work focused on heavy baryonic DM, here we extend the construction to several light dark-quark flavors and investigate the phenomenology of the resulting dark pions as the DM candidates. Our aim is to study their thermal production and the experimental signatures arising from their connection to the neutrino sector.

Dark pion DM models associated with approximate dark-flavor chiral symmetries have been extensively studied in the context of strongly interacting massive particles (SIMPs), with their relic abundance controlled by 3-to-2 reactions within the dark sector. Additional mediators and alternative mechanisms for determining the pion abundance have also been explored in a variety of related scenarios~\cite{Hochberg:2014kqa,Hochberg:2015vrg, Kuflik:2015isi,Bernal:2015bla, Bernal:2015xba, Bernal:2015ova, Choi:2015bya, Choi:2016hid,Soni:2016gzf, Kamada:2016ois,  Kamada:2017tsq,Bernal:2017mqb,Cline:2017tka, Choi:2017mkk,  Kuflik:2017iqs, Heikinheimo:2018esa, Choi:2018iit,Hochberg:2018rjs, Bernal:2019uqr, Choi:2019zeb,  Katz:2020ywn, Smirnov:2020zwf, Xing:2021pkb, Braat:2023fhn,Garani:2021zrr, Dey:2016qgf, Bernreuther:2023kcg,Chu:2024rrv,Chu:2025hga,Garcia-Cely:2024ivo,Garcia-Cely:2025flv,Davighi:2024zip,Davighi:2025awm,Fiorentino:2026dea,Figueroa:2026bmi,Chu:2026rma}. Recently, the possibility of exploring confining sectors through resonant scattering on the cosmic neutrino background has been investigated in~\cite{Cruz:2026dwa}.

In our construction, obtaining the observed pion abundance through thermal freeze-out is not automatic. When the sterile neutrinos are heavy, DM annihilation into active neutrinos is suppressed by their small  masses. Increasing this rate requires a small dark pion decay constant, which also enhances elastic pion scattering. In the mass range of interest, the relic-abundance requirement is generically incompatible with bounds on DM self-interactions. This tension motivates additional depletion channels that can reconcile DM thermal production with the existing bounds. 

A further motivation for light DM comes from the mild excess of electron-antineutrino events around $22$ MeV reported by the Super-Kamiokande (SK) collaboration, usually interpreted in terms of the diffuse supernova neutrino background~\cite{Suzuki:2019jby,Super-Kamiokande:2021the,Super-Kamiokande:2024kcb,Super-Kamiokande:2025sxh}. Ref.~\cite{Granelli:2026bem} investigated a possible interpretation in terms of MeV-scale DM annihilating predominantly into neutrinos and highlighted the prospects for testing this hypothesis at the JUNO detector~\cite{JUNO:2015zny,JUNO:2022lpc}. This raises the question of whether a concrete model of composite DM can produce the required neutrino signal while remaining compatible with its cosmological abundance.

We address these issues by proposing two alternative model-building strategies. In the first scenario, we introduce a light pseudo-scalar mediator, which enables dark-pion semi-annihilations into neutrinos and antineutrinos. We compute neutrino spectra, perform a dedicated fit to SK data, and assess the sensitivity of JUNO to DM semi-annihilations. We also compare the present-day rates with those relevant for freeze-out, including thermal effects near kinematic thresholds. 
In the second scenario, we reconcile DM thermal production with the existing limits, by lowering the scale of the  sterile-neutrinos below the dark-pion mass. Annihilations into sterile neutrinos then provide an efficient depletion channel. While this scenario is not intended to explain the SK excess, it offers complementary tests through sterile-neutrino searches at future accelerator facilities and potentially through gamma-ray and neutrino searches for DM annihilation products.

The paper is organized as follows: we introduce our model in Section~\ref{sec:simplemodel}. We discuss the SK excess and the prospects at the JUNO detector in Section~\ref{sec:phenoheavy}, while we explore the consequences of a low-scale sterile in Section~\ref{sec:phenolight}. We draw our conclusions in Section~\ref{sec:conclusions}.


\section{Neutrino portal to confining dark sectors through inverse seesaw}\label{sec:simplemodel}

The inverse seesaw mechanism accommodates small neutrino masses without requiring tiny Yukawa couplings, by invoking a small lepton-number-violating parameter and allows sterile neutrinos at experimentally accessible scales. However, the origin of the small lepton-number-violating parameter is often left unexplained. This motivates a construction in which such smallness is generated dynamically~\cite{Ma:2009gu,Ahriche:2016acx,Coito:2022kif}. Recently, Ref.~\cite{Berbig:2024uwm} proposed an alternative based on strongly coupled dynamics:
the neutrino sector is connected to a new confining sector, which can generate mass scales dynamically while also providing composite DM candidates. 

We consider a simple model as an illustrative benchmark. 
We choose the confining gauge group to be $\SU(3)_{\rm D}$, to draw analogies with ordinary QCD. We introduce $N_f$ flavors of degenerate vector-like dark quarks $q$, transforming in the fundamental representation of $\SU(3)_{\rm D}$ and a real scalar field $\sigma$, singlet of $\SU(3)_{\rm D}$. 
We impose flavor universality, so that the dark quarks are all degenerate and couple to the scalar $\sigma$ with the same strength. This is equivalent to postulating a global $\SU(N_f)_V$ symmetry acting on the dark quark flavors as $q\to U\,q$.
We further introduce three generations of vector-like gauge singlet neutrinos $(N_L,N_R)$. All the new degrees of freedom are uncharged under the SM gauge group. Following closely the model of Ref.~\cite{Berbig:2024uwm}, we introduce a $\mathcal{Z}_4$ symmetry acting with the following charges: $-1$ on $\sigma$, $-i$ on $q_L$, $+i$ on $\{q_R,N_L,N_R,L,e_R\}$ and +1 on the other SM fields.

The relevant terms in the Lagrangian read 

\begin{equation}\label{eq:FundamentalLagrangian}
\begin{aligned}
    \mathcal{L}\subset&-y_q\,\sigma\bar{q}q-\frac{1}2{m_\sigma^2}\sigma^2\\
    &-\textbf{y}_\nu^{ij} \bar{L}^i\tilde{H}N_R^j-\textbf{M}_D^{i j}\overline{N_R^i} N_L^j-\frac{1}{2}\bm{y}_{N\sigma}^{ij} \sigma\overline{N^{c\,i}_L}N_L^j+h.c.\,,
    \end{aligned}
\end{equation}
where dark-quark flavor indices have been suppressed for simplicity.
A few comments are in order: $L$ is the SM doublet field, while $\tilde{H}=i\sigma_2 H^*$ is the dual SM Higgs field; $y_q$ is a real number; the Yukawa $\textbf{y}_\nu$ and the Dirac mass term $\textbf{M}_D$ are complex $3\times 3$ matrices, while $\textbf{y}_{N\sigma}$ is a complex symmetric $3\times 3$ matrix.
While generically allowed, the Majorana term $\sigma \overline{N_R^c}N_R$  does not introduce new qualitative effects for our discussion and we ignore it in the following. Notice that the $L HN_L$ term is forbidden by the $\mathcal{Z}_4$ symmetry. This leads to a viable realization of the inverse seesaw mechanism, as we will discuss in the following.

The $\SU(3)_{\rm D}$ gauge interactions confine at some energy scale $\Lambda$. In analogy with ordinary  QCD, it is believed  that gauge confinement is accompanied by the generation of a  fermion condensate, $\langle{\bar{q}q\rangle}\simeq \Lambda^3$. The condensate generates a linear term in the scalar potential which induces a vacuum expectation value (vev) for the scalar field $\langle\sigma\rangle\sim y_q\Lambda^3/m_\sigma^2$. This simultaneously generates the Majorana mass parameter and the dark quark mass as 

\begin{equation}
\bm{\mu}=\bm{y}_{N\sigma}\langle\sigma\rangle\,,\qquad m_q=y_q\langle\sigma\rangle\,.
\end{equation}
Notice that this dynamical mechanism has the following advantages: the $L H N_L$ term is forbidden by the discrete symmetry; a small lepton-number violating parameter $\mu$ term is generated simply by imposing a hierarchy of scales $\Lambda\ll m_\sigma$; the confining sector also provides composite DM candidates, linking neutrino mass generation and DM within a common framework. While Ref.~\cite{Berbig:2024uwm} focused on a baryonic DM candidate with a mass of $\sim100$ TeV, here we investigate the phenomenology of light pseudo-Nambu-Goldstone composite DM.

\subsection{The dark pion sector}

If $m_q\ll \Lambda$, the Lagrangian exhibits an approximate $\SU(N_f)_L \otimes \SU(N_f)_R$ symmetry, corresponding to independent chiral transformations of the left- and right-handed fermion components in flavor space.
The fermion condensate spontaneously breaks the global symmetry group of the Lagrangian to the diagonal subgroup $\SU(N_f)_V$, corresponding to those transformations acting in the same way on left and right chiral components. This gives rise to a spectrum of $N_\pi=N_f^2-1$ pseudo-Goldstone bosons $\pi^a$, from now on referred to as dark pions. At sufficiently low energies, their interactions are described by Chiral Perturbation Theory (ChPT) as (for a review of standard ChPT techniques, see~\cite{Pich:1995bw,Scherer:2002tk})

\begin{equation}\label{eq:ChiPTLagrangian}
    \mathcal{L}_{\rm eff}= \frac{f_\pi^2}{4}{\rm Tr}[\partial_\mu U^\dagger \partial^\mu U]+\frac{B_0m_qf_\pi^2}{2}{\rm Tr}[U+U^\dagger]+ y_q\tilde{\sigma}\frac{B_0f_\pi^2}{2}{\rm Tr}[U+U^\dagger]\,,
\end{equation}
where $U=\exp[i\pi_a\sigma_a/f_\pi]$ is the pion field matrix, with $\sigma_a$ being the generators of $\SU(N_f)$ normalized as ${\rm Tr}[\sigma_a\sigma_b]=2\delta_{ab}$. The dark pion decay constant is expected to scale as $\Lambda\sim c_ff_\pi$, where in ordinary QCD, $c_{f}^{\rm QCD}\simeq 2.7$~\cite{Murayama:2026ioh}, while
the fermion condensate is parametrized as $\langle\bar{q}q\rangle\equiv-B_0f_\pi^2$.  The scalar excitation has been defined as $\sigma(x)=\langle\sigma\rangle+\tilde{\sigma}(x)$.
Notice that Ref.~\cite{Berbig:2024uwm} considered only one light flavor of dark quarks, so that the light pseudo-Goldstone bosons were absent from the spectrum. Furthermore, 
the dark confinement scale was assumed to lie around $\sim 100$ TeV, and the DM was identified with the lightest dark baryon. Here, in analogy with the discussion in the previous section, we consider $N_f\geq2$ light flavors and we focus on lighter scales $\Lambda\lesssim 10$ GeV, so that the dark pions emerge as the DM candidates. Finally, the scalar $\tilde{\sigma}$ is assumed to be heavy, with mass $m_\sigma\gg\Lambda$ and can be safely integrated out from the low-energy dynamics.

Expanding the Lagrangian of Eq.~\eqref{eq:ChiPTLagrangian} we get 

\begin{equation}\label{eq:chiral1}
    \mathcal{L}_{\rm eff}\subset -\frac{1}{2}m_\pi^2\pi^2-B_0y_q\tilde{\sigma}\pi^2+\dots\,,
\end{equation}
where $\pi^2=\pi_a\pi_a$, while the dark pion mass is generated dynamically as $m_\pi^2=2B_0m_q$.
In light of their pseudo-Goldstone nature, the dark pions are the lightest states of the dark confining sector, with a mass scaling $m_\pi\sim \sqrt{m_q\Lambda}$, while additional resonances (such as baryons, glueballs and vector mesons) have masses comparable to $\Lambda$.
The unbroken $\SU(N_f)_V$ guarantees the stability of the dark pions, so that they are natural DM candidates\footnote{Note that the heavy baryons are also stable, in light of an accidental $U(1)$ symmetry, corresponding to a re-phasing of the dark quark field. However, for the mass scales considered in this work, $\Lambda\lesssim\mathcal{O}(10)$ GeV, and neglecting any primordial asymmetry, their contribution to the DM abundance is negligible.}. Notice that the DM mass is generated by the same dynamics responsible for the small active neutrino masses, providing a strong connection between the two sectors.

\subsection{The neutrino sector and the inverse seesaw}

We now discuss the neutrino sector. 
We work under the assumption that
\begin{equation}
    \mu\ll M_D\,\,\,\,\,\text{and}\,\,\,\,\,y_\nu v_H\ll M_D\,,
\end{equation}
where $\langle{H\rangle}=v_H=174$ GeV indicates the SM Higgs vev, while,  $\mu\,,y_\nu\,,M_D$ represent the typical order of magnitude of the elements of the corresponding matrices. Within this assumption, the active neutrino mass matrix is obtained as~\cite{CentellesChulia:2020dfh}

\begin{equation}\label{eq:numass}
    \textbf{m}_\nu\simeq v_H^2 \textbf{y}_\nu^* (\textbf{M}_D^T)^{-1}\bm{\mu}\, \textbf{M}_D^{-1}\textbf{y}_\nu^\dagger\,,
\end{equation}
so that active neutrino masses are parametrically suppressed by the small Majorana mass, $m_\nu\sim \mu\,(y_\nu v_H/M_D)^2$. The two chiralities of heavier neutrinos, from now on referred to as the sterile neutrinos, form a pseudo-Dirac pair $(N_1,N_2)$ with masses $M_{N_{1,2}}\simeq M_D\pm\mu/2$ (one pair for each of the three generations).
Known as inverse seesaw~\cite{Mohapatra:1986bd,Gonzalez-Garcia:1988okv} (see also~\cite{Cai:2017jrq}), this mechanism has been proposed as a low-scale variant of the canonical seesaw. Indeed, for $y_\nu\sim \mathcal{O}(1)$ and $\mu\sim \mathcal{O}(\text{keV})$, the desired values of active neutrino masses, $m_\nu\sim 0.05$ eV, are obtained for sterile neutrinos of mass $M_{N_{1,2}}\simeq M_D\sim 1-10$ TeV. The inverse seesaw mechanism typically predicts relatively large mixings $|V_{\alpha N_k}|^2\sim m_\nu/\mu$ between active neutrinos of flavor $\alpha=e,\mu,\tau$ and sterile mass eigenstates $N_k$. Finally the interaction of the scalar $\tilde\sigma$ with active neutrinos is obtained by rescaling the corresponding coupling to sterile species, $y_{N\sigma}$ by the active-sterile mixing angle. Given that $y_{N\sigma}\sim \mu/\langle\sigma\rangle$, and $|V_{\alpha N_k}|^2\sim m_\nu/\mu$, the $\tilde\sigma\nu\nu$ interaction turns out to be proportional to the active neutrino mass matrix.

\subsection{Dark pion annihilation into active neutrinos}

When both the scalar $\tilde\sigma$ and the sterile neutrinos are heavy compared with the dark pions, we can integrate them out from the low-energy theory. Thus, the low-energy interactions among dark pions and active neutrinos are described schematically by

 \begin{equation}
     \mathcal{L}_{\pi\pi\nu\nu} \simeq\sum_{i}\frac{m_{\nu_i}}{2N_{f} f_{\pi}^{2}}\overline{\nu_{i}}\nu_{i}\pi^{2}\,,
    \end{equation}
where $\nu_i$ denotes an active-neutrino mass eigenstate. Once the neutrino masses and $N_f$ are specified, the interaction strength is determined entirely by $f_\pi$. Such interaction gives rise to DM annihilations into active neutrinos, $\pi\pi\to \nu\nu$, whose corresponding cross section scales as $\sigma v\sim m_\nu^2/f_\pi^4$, and is therefore strongly suppressed by the small neutrino masses. For an initially thermal population of dark pions, reproducing the observed cosmological DM abundance measured by the Planck collaboration~\cite{Planck:2018vyg} through thermal freeze-out, driven by the $\pi\pi\to\nu\nu$ annihilations, requires a sufficiently large annihilation rate. This corresponds to $ f_\pi\sim (M_{\rm Pl}T_{\rm eq}\sum m_\nu^2)^{1/4}\sim 100\text{ keV}-1\text{ MeV}$. More conservatively, requiring only that the thermal pion abundance does not exceed the observed DM density leads to an equivalent upper bound of $f_\pi\lesssim 1$ MeV.

Despite being very predictive, this simple setup shows a clear tension. Indeed, the scale $f_\pi$ also controls elastic pion scattering, whose cross section is (taking into account the standard chiral computation and the additional contribution generated by the exchange of $\sigma$)

\begin{equation}
    \sigma_{\pi\pi}=\frac{3(N_f^2+4)}{64\pi(N_f^2-1)}\frac{m_\pi^2}{f_\pi^4}\overset{N_f=3}{=}\frac{117m_\pi^2}{1536\pi f_\pi^4}\,.
\end{equation}
Constraints from the Bullet Cluster and other galaxy clusters~\cite{Harvey:2015hha,Bondarenko:2017rfu,Harvey:2018uwf,Sagunski:2020spe,DES:2023bzs} require $\sigma_{\pi\pi}/m_\pi\lesssim0.5\text{ cm}^2/\text{g}$. 
For dark pion masses $m_\pi\gtrsim10\,\mathrm{MeV}$, these bounds require $f_\pi$ of order tens of MeV or larger. Lighter thermal relics coupled to neutrinos can instead be constrained by entropy transfer during neutrino decoupling and the resulting change in $N_{\rm eff}$.

The requirements from the relic abundance and self-interactions therefore select incompatible ranges of $f_\pi$. Lowering $f_\pi$ enough to obtain efficient annihilation violates the self-interaction bounds, whereas increasing it to satisfy those bounds suppresses annihilation and leaves an overabundant population of DM, excluded by cosmological data. 

This tension motivates model-building strategies for additional depletion channels that avoid the strong suppression by active neutrino masses. We explore two possibilities. In Section~\ref{sec:phenoheavy}, we retain heavy sterile neutrinos and introduce a light pseudo-scalar $a$, enabling dark-pion semi-annihilation. We investigate whether this mechanism can reproduce the thermal relic abundance while satisfying constraints from DM self-interactions.
Interestingly, this possibility typically selects DM masses in the $\sim 10$ MeV mass range, which can be tested at the current generation of neutrino telescopes, such as SK and JUNO. We will explore the corresponding phenomenology with a particular focus on the $22$ MeV antineutrino excess reported by the SK collaboration. 

Alternatively, in Section~\ref{sec:phenolight} we retain the original particle content of Eq.~\eqref{eq:FundamentalLagrangian} and lower the sterile-neutrino masses below the dark-pion mass, opening the annihilation channel $\pi\pi\to NN$. This provides a new possibility for thermal freeze-out, with complementary prospects for tests at future accelerator facilities.

\section{Light Dark Matter and its phenomenology at neutrino telescopes}\label{sec:phenoheavy}

\subsection{Dark pion semi-annihilations into active neutrinos}

\paragraph{A new light degree of freedom.} We extend the Lagrangian of Eq.~\eqref{eq:FundamentalLagrangian} with a light degree of freedom, namely a pseudo-scalar $a$, singlet of $\SU(3)_D$, with mass $m_a\ll\Lambda$. We assign it the same charge of $\sigma$ under the $\mathcal{Z}_4$ symmetry, namely $a\to-a$. Such a light state could arise as a pseudo-Nambu-Goldstone boson associated with an additional approximate global symmetry. It could also be embedded, together with the scalar $\sigma$, into a single complex field $\varphi=\sigma+i\,a$.  Although such an embedding is possible and relatively minimal, maintaining the hierarchy $m_a\ll m_\sigma$ requires fine-tuning. For the phenomenological analysis below, we treat $\sigma$ and $a$ as independent fields, keeping in mind a possible common origin (a dedicated UV model able to explain the fine-tuning is left as an interesting future direction).

The Lagrangian is extended with the extra terms

\begin{equation}\label{eq:deltalag}
    \Delta \mathcal{L}=-y'_q\,a\,\bar{q}i\gamma_5q-\frac{1}2{m_a^2}a^2-i\,\frac{\textbf{y}_{Na}^{ij}}{2}a\overline{N^{c\,i}_L}N_L^j+{\rm h.c.}\,,
\end{equation}
where $\bm{y}_{Na}$ is a symmetric complex $3\times 3$ matrix \footnote{In the absence of an additional protective symmetry, loops involving the heavy sterile neutrinos can generically induce threshold corrections to the pseudo-scalar mass. We therefore treat $m_a$ as a renormalized input parameter and accept the fine-tuning required to keep it in the MeV range. An ultraviolet completion protecting this hierarchy is left for future work.}. Notice that the operator $\bar{q}\gamma_5q$ does not develop a vacuum condensate, unlike the $\bar{q}q$ counterpart. As a result, the pseudo-scalar $a$ does not contribute to the generation of active neutrino masses, or conversely, the interactions among $a$ and active neutrinos are not suppressed by their tiny masses. This differentiates the dynamics of the pseudo-scalar with respect to that of the scalar $\sigma$.

The low-energy interactions of dark pions are augmented with the new operator

\begin{equation}\label{eq:chiral2}
    \mathcal{L}_{\rm eff}^{\pi a}= -i\,ay'_q\frac{B_0f_\pi^2}{2}{\rm Tr}[U-U^\dagger]=-\frac{y'_qB_0}{3f_\pi}d_{abc}\,a\,\pi_a\pi_b\pi_c\,,
\end{equation}
where $d_{abc}$ denote the symmetric structure constants of $\SU(N_f)$. Notice that, in this case, one needs $N_f\geq 3$ since for $N_f=2$ the symmetric constants vanish.

\vspace{0.3cm}

\paragraph{Choice of the parameters.}In order to emphasize the physics case, we restrict our analysis to a simplified scenario which captures the relevant physical implications, avoiding unnecessary complications related to the flavor structure. Working in a basis in which the charged lepton masses are diagonal, we consider the following choice of mass and Yukawa matrices

\begin{equation}\label{eq:benchmasses}
    \textbf{M}_D=M_D\mathbf{1}_3,\qquad \textbf{y}_{Na}=y_{Na}\mathbf{1}_3,\qquad \bm{\mu}=\textbf{m}_\nu^{\rm diag}/\theta^2={\rm diag}(m_{\nu_i}/\theta^2),\qquad v\,\mathbf{y}_\nu=\theta M_DU_{\rm PMNS}\,,
\end{equation}
with the mixing angle
\begin{equation}
    \theta^2\simeq \sum_{k=1}^6|V_{\alpha N_k}|^2\,\qquad \text{for each}\,\alpha=e,\mu,\tau\,,
\end{equation}
where the sum is performed over the six sterile neutrinos.
We also assume real $y_{Na}$, and neglect any effects of CP-violating couplings for $a$.
Under these assumptions, the interactions between active neutrinos mass eigenstates $\nu_i$ and $a$ read
\begin{equation}\label{eq:LaganuN}
    \mathcal{L}_{a\nu\nu}\simeq \frac{y_{Na}\theta^2}{2}\,a\sum_i\nubar_ii\gamma_5\nu_i\,,
\end{equation}
while $\textbf{m}_\nu=U_{\rm PMNS}^*\mathbf{m}_\nu^{\rm diag}U^\dagger_{\rm PMNS}\,.$
More general flavor structures can generate non-universal and off-diagonal couplings. For generic textures without strong hierarchies, special alignments, or cancellations, inclusive rates involving the $a\nu\nu$ vertex, summed over light-neutrino species, are expected to differ from our benchmark only by factors of order unity, at fixed overall Yukawa and active-sterile mixing scales. To avoid unnecessary and tedious complications, and keep the equations as simple as possible, we stick to our simple choice in Eq.~\eqref{eq:benchmasses} as an illustrative benchmark, leaving a detailed computation for completely generic matrices for future work.
Notice that, if we work in the weak gauge basis, we can simply replace $U_{\rm PMNS}$ with $U_l U_{\rm PMNS}$ in Eq.~\eqref{eq:benchmasses}, with $U_l$ being the charged lepton rotation matrix. 

\vspace{0.5cm}

\paragraph{DM semi-annihilations and neutrino spectra.}The interactions in Eqs.~\eqref{eq:chiral2} and~\eqref{eq:LaganuN} give rise to DM semi-annihilation processes, mediated by the pseudo-scalar particle $a$, producing neutrinos as final states.
We focus on sterile neutrino and DM masses in the ballpark $M_D\sim1-100\text{ TeV}\gg m_\pi\sim10-100$ MeV, so that the heavy steriles can be integrated out.  For future convenience, we define the mass splitting $\Delta$ as

\begin{equation}
    m_a\equiv m_\pi(1+\Delta)\,,\qquad \text{with}\,\,\Delta>-1\,.
\end{equation}
Notice that $\Delta$ can be positive (heavier $a$) or negative (lighter $a$).

To simplify the notation we define $\lambda_1=y'_qB_0/3f_\pi\simeq y'_qc_f^3/3$ and $\lambda_2= y_{Na}\theta^2/2$. Then, the relevant interactions among DM, $a$ and active neutrinos read

\begin{equation}\label{eq:pinulag}
    \mathcal{L}_{\pi\nu}=\lambda_1 d_{abc}\, a\, \pi_a\,\pi_b\,\pi_c+\lambda_2\,a\,\bar{\nu}\,i\,\gamma_5 \nu\,.
\end{equation}
Depending on the hierarchy of scales, we distinguish two possible scenarios: $i)$ if $m_a<m_\pi$, corresponding to $-1<\Delta<0$, DM particles undergo semi-annihilations $\pi\pi\to\pi a$, followed by
decays to active neutrinos $a\to\nu\nu\,(\nubar\nubar)$;
$ii)$ if $m_a>m_\pi$, corresponding to $\Delta>0$, DM undergo direct semi-annihilations to neutrinos or antineutrinos, $\pi\pi\to\pi\nu\nu\,(\pi\nubar\nubar)$, mediated by the exchange of an off-shell $a$.  Notice that in both cases the final state contains two neutrinos or two antineutrinos, both with branching ratio 1/2. 

\begin{figure}
    \centering
    \includegraphics[width=0.9\linewidth]{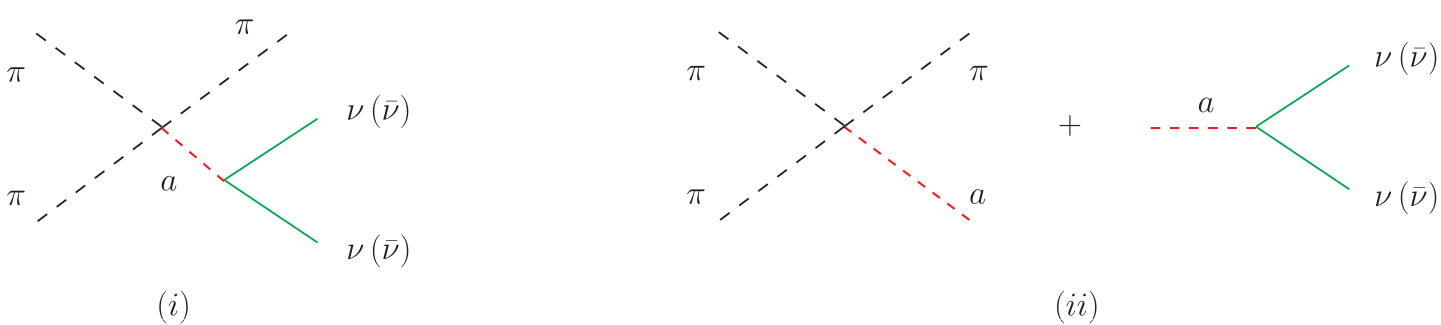}
    \caption{DM semi-annihilations $\pi \pi\to\pi \,\nu\nu\, (\pi\,\nubar\nubar)$, mediated by the exchange of off-shell (left) or on-shell (right) pseudo-scalar $a$.}
    \label{fig:diag}
\end{figure}

\begin{figure}
    \centering
    \includegraphics[width=0.49\linewidth]{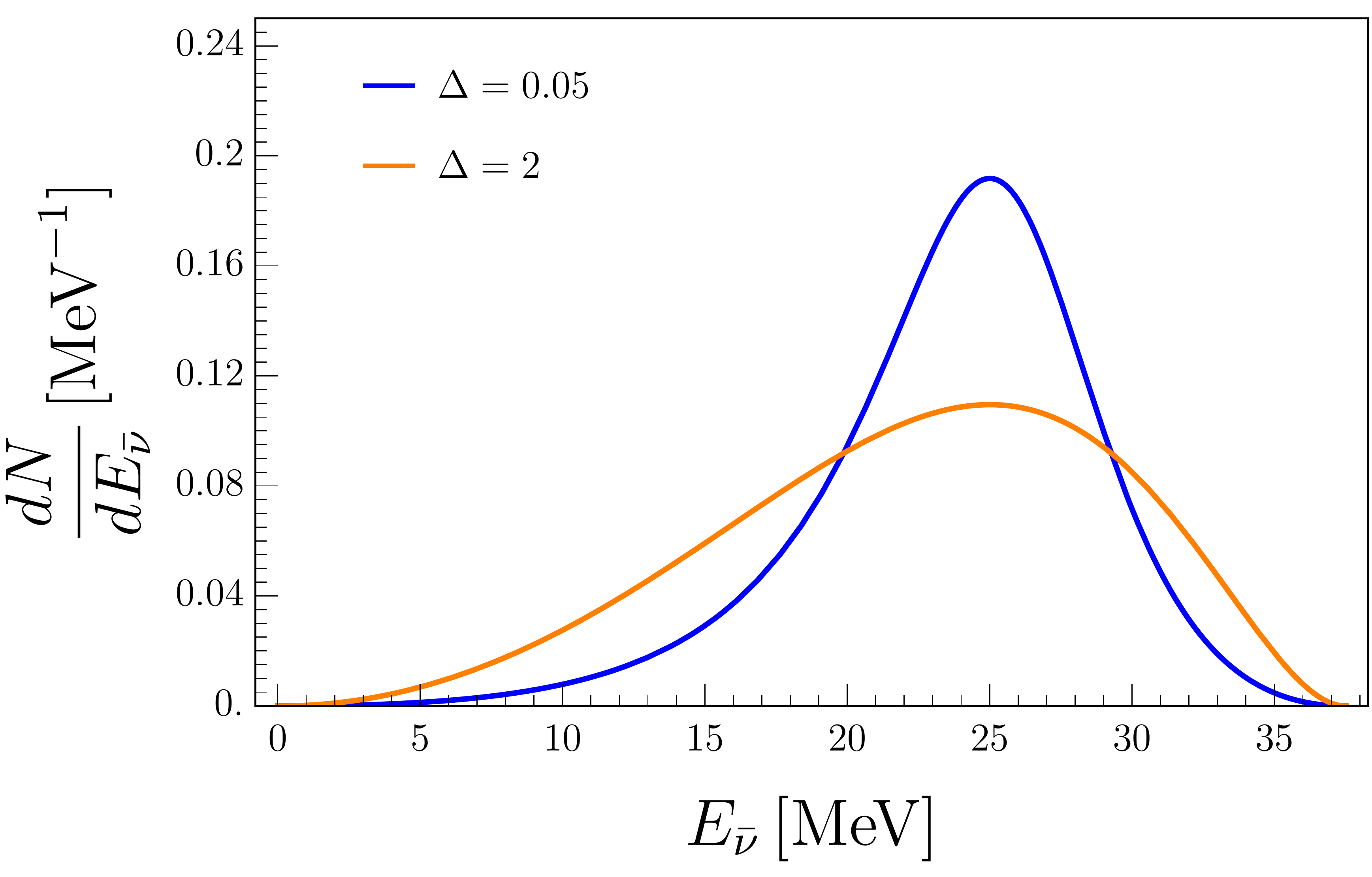}  \includegraphics[width=0.49\linewidth]{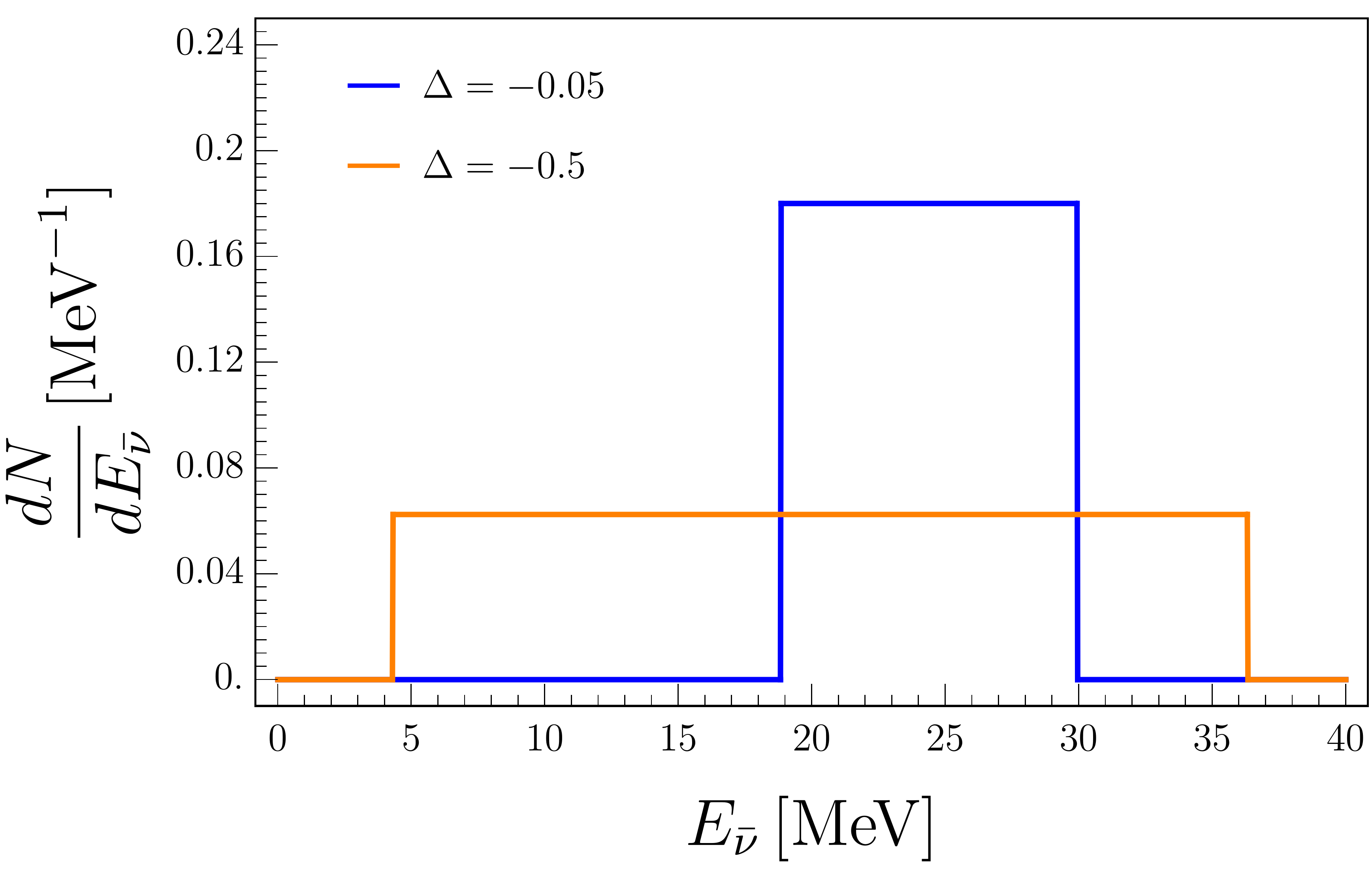}
    \caption{Neutrino energy spectrum for direct semi-annihilations (left) and semi-annihilations followed by decays (right) for different values of mass splitting $\Delta$ and for $m_\pi=50$ MeV.}
    \label{fig:dNdE}
\end{figure}

In the first scenario, DM semi-annihilations proceed with cross section

\begin{equation}\label{eq:22}
    \sigma_{\pi a} v\simeq \frac{45 \lambda_1^2 \sqrt{1-\frac{10\,m_a^2}{9\,m_\pi^2}+\frac{m_a^4}{9m_\pi^4}}}{256\pi\, m_{\pi}^2}\,,
\end{equation}
where $v$ is the relative velocity between the two initial pions.
The pseudo-scalar is produced with energy $E_a=m_\pi[1+\Delta(\Delta+2)/4]$, and subsequently decays to two neutrinos or two antineutrinos with total rate
$\Gamma_a={3\lambda_2^2m_a}/{4\pi}$.
In the rest frame of the pseudo-scalar, both neutrinos are emitted with energy $E_{\nubar}^*=m_a/2$. Performing a boost into the center-of-mass frame of the annihilating DM particles, the neutrino energy is distributed uniformly between $E_{\nubar}^\pm=\gamma_a^{-1} (m_a/2)[1\pm \beta_a]^{-1}$, where $\beta_a,\gamma_a$ are the appropriate relativistic boost factors. The number of antineutrinos emitted per unit energy per annihilation is given by a box spectrum with extreme point $E_{\nubar}^\pm$ and height 

\begin{equation}
    \frac{dN_{\bar{\nu}}}{dE_{\bar{\nu}}}=\frac{8}{m_\pi}[\Delta(\Delta+4)(\Delta^2-4)]^{-1/2}\,.
\end{equation} 
On the other hand, in the second scenario, the total semi-annihilation cross section is defined as 

\begin{equation}
    \sigma_{23}\,v=\sigma_{\pi\nu\nu}\,v+\sigma_{\pi\nubar\nubar}\,v=2\sigma_{\pi\nubar\nubar}\,v
\end{equation}
and is given by (see Appendix~\ref{app:23cross} for details)

\begin{equation}\label{eq:23}
    \sigma_{23}\,v= \frac{45\lambda_1^2\lambda_2^2}{64\pi^3\,s^2}\int_0^{(\sqrt{s}-m_\pi)^2} dq^2 \frac{q^2\sqrt{\lambda_{\rm Kal}(s,m_\pi^2,q^2)}}{(q^2-m_a^2)^2} \overset{m_\pi\ll m_a}{\simeq} \frac{45\lambda_1^2\lambda^2_2\,m_\pi^2}{2048\pi^3m_a^4}(57-40\ln4)\,,
\end{equation}
where $\lambda_{\rm Kal}$ is the K\"{a}ll\'{e}n function.
From Eq.~\eqref{eq:23}, we can compute the number of antineutrinos emitted per unit of energy per annihilation as 

\begin{align}
    \frac{dN_{\nubar}}{dE_{\nubar}}=\frac{2}{\sigma_{\pi\nubar\nubar}\,v}\frac{d\sigma_{\pi\nubar\nubar} v}{dE_{\nubar}}\,.
\end{align}
where the factor of 2 takes into account that 2 antineutrinos are produced in a semi-annihilation process.
The spectra for both scenarios are shown in Figure~\ref{fig:dNdE}.
Throughout the paper we adopt $c_f=c_f^{\rm QCD}\simeq 2.7$ in all our numerical calculations.

Notice that the cross sections in Eq.~\eqref{eq:22} and Eq.~\eqref{eq:23}, and the corresponding neutrino spectra, are evaluated in the non-relativistic limit $s\simeq4m_\pi^2$ and assuming small DM velocities, which is appropriate to describe DM semi-annihilations occurring at present time in galactic halos, for the mass splittings considered here. 
When describing the thermal evolution of DM, instead, the role of kinetic energy can become important. In particular, when $\Delta$ is positive but sufficiently small, the high-energy tail of the pion distribution can make the channel $\pi\pi\to\pi a$ kinematically accessible, even though it is forbidden at zero velocity. We will discuss the corresponding implications when analyzing the DM cosmological production.

\vspace{0.2cm}

\subsection{DM indirect searches at neutrino telescopes}

DM semi-annihilations in galactic halos generate neutrino fluxes, potentially observable by the current generation of neutrino experiments. We focus on the neutrino signal from DM semi-annihilations in the Milky Way halo. The differential diffuse electron antineutrino flux is given by

\begin{equation}\label{eq:nuflux}
    \frac{d\Phi_{\nubar}}{dE_{\nubar}}=\frac{1}{2}\frac{\langle{\sigma v\rangle}}{8\pi m_\pi^2}\frac{1}{3}\frac{dN_{\nubar}}{dE_{\nubar}}J\,.
\end{equation}
Here the semi-annihilation cross section $\sigma v$ corresponds to either Eq.~\eqref{eq:22}, if $\Delta<0$, or Eq.~\eqref{eq:23}, if $\Delta>0$, while $dN_{\nubar}/{dE_{\nubar}}$ is the corresponding neutrino spectrum.
The symbol $\langle{\rangle}$ indicates the average  over the present DM velocity distribution, while the factor 1/2 takes into account that only a half of the semi-annihilation processes produce antineutrinos, and the factor of 1/3 follows from the equal production of the three light-neutrino mass eigenstates in our benchmark and the unitarity of the PMNS matrix. 
For non-universal couplings, after averaging over propagation phases on astrophysical baselines, the factor 1/3 would be replaced by
$
f_e=\sum_{i=1}^{3}p_i\,|(U_{\rm PMNS})_{ei}|^2$
where $p_i$ denotes the fraction of the produced antineutrino flux in mass eigenstate $i$, and satisfies $\sum_i p_i=1$. For generic flavor textures without strong hierarchies, special alignments, or cancellations, this modifies the electron antineutrino flux normalization only by factors of order unity.

The $J$-factor is the line-of-sight integral of the square of the local DM density 

\begin{equation}
    J=\int_{\Delta\Omega}d\Omega\int_0^{\ell_{\rm max}}d\ell\rho_{\rm DM}^2(r(\ell,\cos\alpha))\,.
\end{equation}
Here $r(\ell,\cos\alpha)=\sqrt{R_0^2+\ell^2-2\ell R_0\cos\alpha}$, where $R_0\simeq 8.277$ kpc is the galactocentric distance~\cite{GRAVITY:2021xju,Abuter:2021yys}, $\alpha\in [0,\pi]$ is the angle between the line-of-sight $\ell$ and the Sun-Galactic Center direction and $\Delta\Omega$ the total solid angle of view, equal to $4\pi$ for full-sky signals. For a spherical halo of radius $R_{\rm halo}$, $\ell_{\rm max}=\sqrt{R_{\rm halo}^2-R_0^2(1-\cos\alpha^2)}+R_0\cos\alpha$, and the integral is marginally dependent on the specific value of $R_{\rm halo}$, if sufficiently large. In Ref.~\cite{Granelli:2026bem} (see also~\cite{Benito:2019ngh,Benito:2020lgu}), the authors fixed $R_{\rm halo}=100$ kpc and adopted the generalized Navarro-Frenk-White DM density profile~\cite{Navarro:1995iw} to fit the Milky Way rotation curve data. Normalizing $\mathcal{J}=J/J_0$, with $J_0=4\pi R_0\rho_0^2$ and $\rho_0\simeq 0.4\,\text{GeV}/\text{cm}^3$, they performed a fit of the Milky Way rotation curve data from stellar tracers in circular orbit up to $20$ kpc. Adopting the likelihood in~\cite{Benito:2019ngh}, they
found viable values of the $J$-factor in the range $4\lesssim \mathcal{J}\lesssim 17$ at $2\sigma$.

\vspace{0.2cm}
Neutrino experiments such as SK~\cite{Suzuki:2019jby} and JUNO~\cite{JUNO:2015zny,JUNO:2022lpc}, as well as the next-generation Hyper-Kamiokande experiment~\cite{Hyper-Kamiokande:2018ofw}, can operate as neutrino telescopes by detecting astrophysical fluxes of electron antineutrinos through
inverse beta decay (IBD) on free protons in the detector, $\bar{\nu}_e+p\to e^++n$. The corresponding signal consists of a prompt component, produced by the energy deposition and subsequent annihilation of the positron, with reconstructed visible energy
$E_{\rm vis}=E_{\nubar}-m_n+m_p+m_e\simeq E_{\nubar}-0.782$ MeV, followed by a delayed component associated with the capture of the produced neutron.
The number of expected events in the $i$-th reconstructed-energy bin is given by

\begin{equation}\label{eq:Nevents}
    N_i=N_{p}T\int _{E_{\rm vis}^i}^{E_{\rm vis}^{i+1}}dE_{\rm vis}\,\epsilon(E_{\rm vis})\,\int_{E_{\nubar}^{\rm min}}^\infty dE_{\nubar} \frac{d\Phi_{\nubar}}{dE_{\nubar}}\,\int_{E_{e}^-(E_{\nubar})}^{E_{e}^+(E_{\nubar})}dE_e\,\frac{d\sigma_{\rm IBD}}{dE_e}\,R(E_e+m_e,E_{\rm vis})\,,
\end{equation}
where $N_p$ is the number of free protons in the fiducial mass of the detector, $T$ is the running time, $\epsilon$ is the efficiency of the detector, $d\sigma_{\rm IBD}/dE_e$ is the differential IBD cross section taken from~\cite{Super-Kamiokande:2008ecj} and $R$ is a Gaussian smearing,

\begin{equation}
    R\left(E,E_{{\rm vis}}\right)=\frac{1}{\sqrt{2\pi}\sigma_{E}}\exp\left[-\frac{1}{2}\left(\frac{E_{{\rm vis}}-E}{\sigma_{E}}\right)^2\right]\,,
\end{equation}
where the energy resolution $\sigma_E$ depends on the experiment. The two kinematic endpoints $E_{e}^\pm$ are obtained imposing energy and momentum conservation at fixed incoming antineutrino energy $E_{\nubar}$ (in the laboratory frame). Notice that the detector dependence enters through the exposure $N_p\times T$, the efficiency $\epsilon$ and the response function $R$.

\subsubsection{SK $\bar{\nu}_e$ excess at $E_\nu\simeq 22$ MeV}

As discussed extensively in Ref.~\cite{Granelli:2026bem}, the SK collaboration has reported an excess of electron antineutrino events around $E_{\nubar}\simeq 22$ MeV, often interpreted as a potential signal of the diffuse supernovae background (DSNB), namely the integrated neutrino flux originating from all past supernovae~\cite{Krauss:1983zn,Beacom:2010kk}. Within this interpretation, the background-only hypothesis is disfavored at up to $\sim 2.3\,\sigma$ when all the SK runs are combined together. The authors of~\cite{Granelli:2026bem} proposed an alternative explanation: the excess could be produced by the annihilations of sub-GeV DM into anti-neutrinos. They do not provide a concrete model, but consider two different phenomenological benchmarks, where an unspecified DM candidate annihilates into (anti)neutrinos either directly, ${\rm DM}\,{\rm DM}\to\nu\bar{\nu}$,  or through some lighter mediator, ${\rm DM}\,{\rm DM}\to\phi\phi\to\nu\bar{\nu}\nu\bar{\nu}$. Treating the DM and mediator masses and the DM annihilation cross section as free parameters, they fit them to the SK data, finding that DM masses in the 20-50 MeV ballpark and cross sections compatible with thermal freeze-out production provide a possible explanation for the SK excess. Finally they discuss the potential of the JUNO detector to test their hypothesis.

The model defined by the Lagrangian in Eq.~\eqref{eq:FundamentalLagrangian}, augmented with the interactions in Eq.~\eqref{eq:deltalag}, provides a concrete realization of the phenomenological framework investigated in Ref.~\cite{Granelli:2026bem}, based on new strongly coupled dynamics. In our setup, neutrino fluxes are generated through DM semi-annihilations, resulting in neutrino spectra that differ from those considered in~\cite{Granelli:2026bem}, which motivated us to perform a dedicated fit to the SK data. 

The SK collaboration has performed six analyses dedicated to DSNB, with more than 6000 days of data taking. They are organized in SK-I to SK-IV (``pure-water era"), SK-IV to SK-VII (``neutron-tagging era") and SK-VI to SK-VII (``Gd-era", in which the SK detector has been loaded with gadolinium(III) sulfate octohydrate, significantly increasing the sensitivity to DSNB). Data from the six periods have been collected from~\cite{PhysRevD.85.052007, PhysRevD.104.122002, Super-Kamiokande:2025sxh}.
The same analysis can be used to study neutrino DM signals in the SK detector, see for instance~\cite{Palomares-Ruiz:2007trf,Olivares-DelCampo:2017feq,Arguelles:2019ouk}.
The number of expected events in the SK detector is given by Eq.~\eqref{eq:Nevents}, where $E_{\nubar}^{\rm min}=17.3$ MeV~\cite{Super-Kamiokande:2025sxh}, while for the energy resolution we adopted $\sigma_E=\text{MeV}\,[0.4\sqrt{E/\text{MeV}}+0.03\,E/\text{MeV}]$~\cite{Palomares-Ruiz:2007trf,Super-Kamiokande:2010tar,Super-Kamiokande:2023jbt} for every SK run except for SK-II, for which we used $\sigma_E=\text{MeV}\,[0.0536+0.52\sqrt{E/\text{MeV}}+0.0458\,E/\text{MeV}]$~\cite{Super-Kamiokande:2008ecj}.
Notice that SK analyses conventionally report the reconstructed kinetic energy of the prompt positron, $E_{\rm rec}\simeq E_{\nubar}+m_p-m_n-m_e\simeq E_{\nubar}-1.8~\mathrm{MeV}$, rather than the total prompt visible energy $E_{\rm vis}$. 
The expression in Eq.~\eqref{eq:Nevents} can be adapted to the SK convention by replacing $E_{\rm vis}\to E_{\rm rec}$, including in the bin boundaries and detection efficiency, and by replacing the detector response according to $ R(E_e+m_e,E_{\rm vis})\longrightarrow R(E_e-m_e,E_{\rm rec})$.

We perform a $\chi^2$ analysis taking into account SK data from all the different periods. The $\chi^2$ function is defined for each SK run as 

\begin{equation}
    \chi^2(m_\pi,\mathcal{J}\langle \sigma v_{\rm rel}\rangle,\alpha,\theta_i)=-2\sum_{i=1}^{N_{\rm bins}}\left[O_i-\alpha N_i-(1+\theta_i)B_i+O_i\log\frac{\alpha N_i+(1+\theta_i)B_i}{O_i}-\frac{\theta_i^2}{2\sigma_i^2}\right]\,,
\end{equation}
where $O_i$ are the observed number of events per energy bin, $B_i$ the background events, while $\alpha$ is the normalization of the signal and $\theta_i$ are nuisance parameters with Gaussian priors of width $\sigma_i$ which parametrize systematic uncertainties in the background.

\begin{figure}[t]
    \centering
    \includegraphics[width=0.5\linewidth]{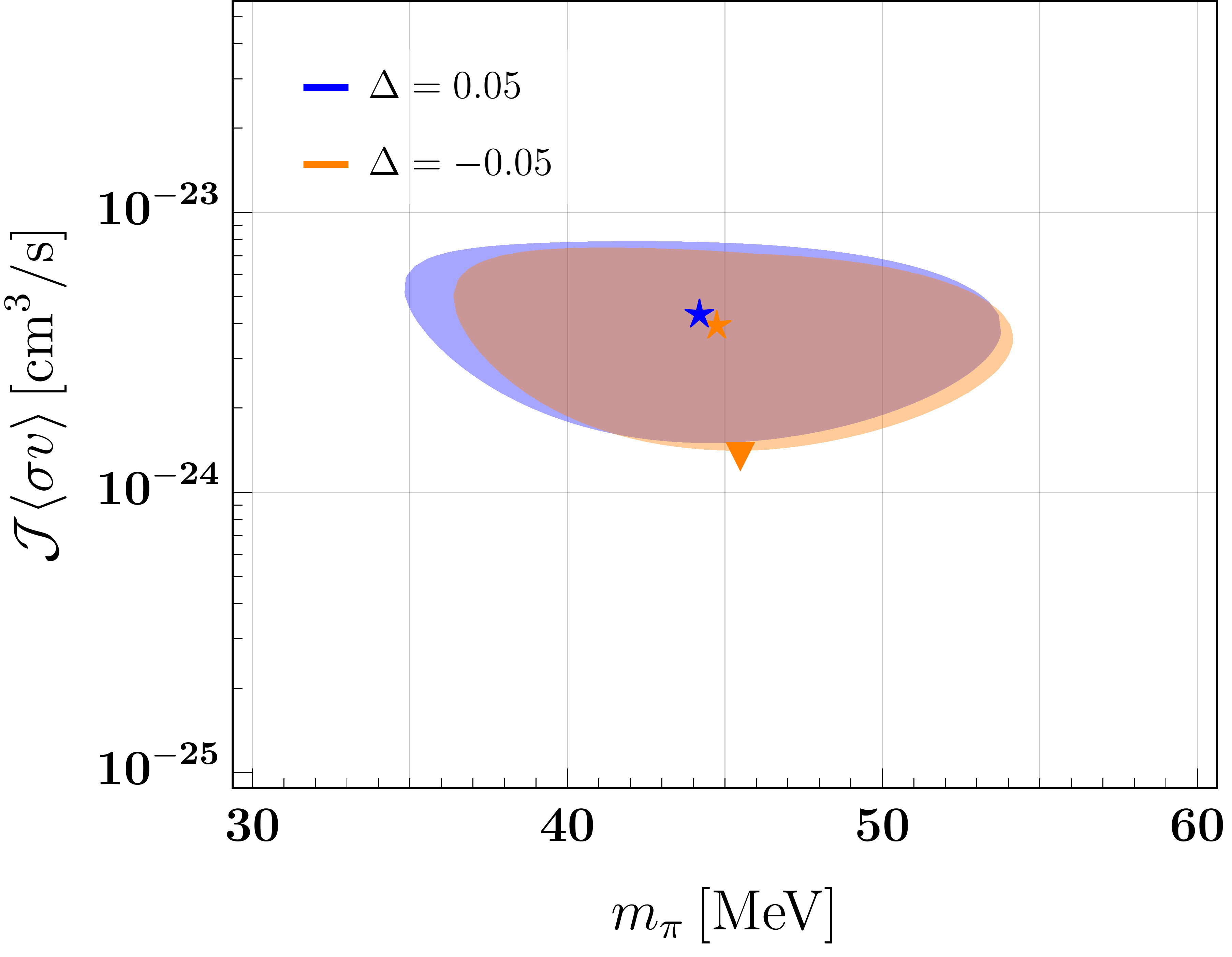}
    
    \caption{$1\sigma$ confidence regions in the $(m_\pi,\mathcal{J}\langle{\sigma v\rangle})$ plane, for two different choices of $\Delta$ and combining all the SK runs discussed in the main text. 
    The two stars mark the best-fit points for each value of $\Delta$, while the triangle marks a point compatible with freeze-out production of DM (see discussion in Section~\ref{sec:thermal})}
    \label{fig:SKmain}
\end{figure}

Since the full likelihood used by the SK Collaboration is not available to us, we construct an approximate likelihood using the background model adopted in the DSNB analyses. For each SK data-taking period, we calibrate the widths $\sigma_i$ on the background nuisance parameters $\theta_i$, defined in each energy bin, to reproduce the corresponding published DSNB likelihood curves (see in particular the  curve in figure 6.16(b) in~\cite{Beauchene:2024DSNB}). We then replace the DSNB signal template with the prediction for DM semi-annihilations and marginalize over the nuisance parameters.

We show in Fig.~\ref{fig:SKmain} the $1\sigma$ confidence regions obtained from our fit for two different values of $\Delta$. 
The two stars correspond to the best-fit points for each choice of $\Delta$. The orange triangle corresponds to a benchmark point compatible with the thermal production of DM, as we will explain in detail in Section~\ref{sec:thermal}. The significance of the DM signal over the null hypothesis is $2.52\sigma$ for the blue star (with $\chi^2=7.8$, for a total of 23 degrees of freedom), and $2.49\sigma$ for the orange star (with $\chi^2=7.9$), while for the orange triangle we get $2.05\sigma$ significance ($\chi^2=10)$, whereas the null hypothesis corresponds to $\chi^2=14.2$.


\subsubsection{Compatibility with thermal freeze-out production}\label{sec:thermal}

In this subsection, we briefly examine whether the present-day semi-annihilation rates needed to explain the SK excess are compatible with the observed DM abundance being produced through standard thermal freeze-out, evaluating the relevant thermally averaged rates.

In the early universe, the interactions among the dark pions, $a$ and neutrinos keep them in thermal equilibrium with the SM thermal bath at a common temperature $T$, as long as $\theta^2\gtrsim 10^{-8}$. The DM number density follows its equilibrium distribution until the universe cools down to $T\sim m_\pi/20$, when the DM annihilation rate drops below the Hubble expansion rate and the dark pions freeze-out leaving a constant relic abundance.
More precisely, the DM relic density is obtained by solving the appropriate Boltzmann equation 
$
\dot{n}_\pi+3Hn_\pi=-\frac{1}{2}\langle\sigma v_{}\rangle\,(n_\pi^2-n_\pi n_{\pi}^{\rm eq})\,,
$
where $n_\pi$ is the dark pion number density, $n_\pi^{\rm eq}$ its equilibrium value, $H$ is the Hubble expansion rate in a radiation-dominated universe and $\langle{\sigma v_{}\rangle}$ is the thermally averaged cross section for the relevant DM semi-annihilation process. 
Notice that for values of DM masses $m_\pi\gtrsim40$ MeV, freeze-out  takes place before neutrino decoupling.

The observed DM cosmological abundance is obtained by matching the DM energy density to the value measured by the Planck collaboration~\cite{Planck:2018vyg}. 
We solved numerically the Boltzmann equation for different combinations of parameters.
Processes other than semi-annihilations might in principle affect the DM evolution, including $3\pi\to2\pi$ processes induced by WZW terms~\cite{Hochberg:2014kqa}, which become relevant whenever $m_\pi\gtrsim 2\pi f_\pi$. However, in most of the region of the parameter space relevant for our phenomenological analysis $f_\pi> m_\pi$ and 3-to-2 processes are negligible. 

\vspace{0.2cm}

For $\Delta<0$, the channel $\pi\pi\to\pi a$ is kinematically accessible both at freeze-out and in present-day galactic halos. While Eq.~\eqref{eq:22} accurately describes the latter, the kinetic energy of the pions can modify the available phase space during freeze-out, particularly for a degenerate spectrum with $|\Delta|\ll1$.
In more detail, the relatively large DM velocities during freeze-out, and the small mass splitting, modify the expression in Eq.~\eqref{eq:22}, which is obtained in the limit of small velocity, $v^2\ll|\Delta|$, valid in present-day DM halos. For the DM thermal evolution we retain the full velocity dependence in the thermally averaged cross section. For the benchmark point marked with an orange triangle in Fig.~\ref{fig:SKmain}, we find that the ratio between the thermally averaged cross section relevant for freeze-out (including full velocity dependence) and the corresponding present-day value (in the small velocity limit) is $\langle{\sigma v\rangle}_{\rm fo}/\langle{\sigma v\rangle}\simeq 1.35$. 
Matching the observed relic abundance fixes $y'_q\simeq4.5\times10^{-6}$,  while $f_\pi$ is left as a free parameter, allowing the model to evade DM self-interaction limits. Accounting for the thermal enhancement and adopting $\mathcal J=17$, the corresponding present-day rate is
$\mathcal J\langle\sigma v\rangle
\simeq1.4\times10^{-24}\,\mathrm{cm^3/s}$,
as shown by the orange triangle. This point lies within the $1\sigma$ region of the SK fit, showing that the same interaction can account for both the thermal relic abundance and a viable explanation for the SK excess. The relic-density requirement fixes $y'_q$, while leaving sufficient freedom in $f_\pi$ to satisfy the DM self-interaction bounds discussed below. The expected signal at the JUNO detector is presented in the next subsection.

The two best-fit points in Fig. 3, marked by the two stars, however, are incompatible with standard thermal production under our assumptions. For the orange star, reconciling the needed present-day rate with the thermal relic abundance would require a very large $\mathcal J$-factor disfavored by current data. For the blue star, with $\Delta=0.05$, the channel $\pi\pi\to\pi a$ is forbidden in halos but becomes accessible through the thermal tail at freeze-out. Despite its Boltzmann suppression, $\exp[-\Delta m_\pi/T]$, it dominates over the three-body process, which carries an additional suppression from $\lambda_2^2$ and phase space. Fixing the relic abundance therefore yields a negligible present-day neutrino signal; conversely, the couplings required by the blue star would deplete the DM below the observed abundance.

These best-fit points could instead be realized through non-thermal production of dark pions after freeze-out. In particular, an alternative production mechanism must be invoked in order to re-populate the DM sector, after the large annihilation cross sections favored by the fit have depleted its thermal abundance. For the nearly degenerate off-shell scenario, another possibility is to introduce a time-dependent mass splitting $\Delta$, larger during freeze-out than today, suppressing the two-body channel in the early universe. Such an evolution could arise by coupling $a$ to an additional scalar and invoking a late phase transition, provided that the mass change does not trigger renewed DM depletion or significant entropy production. A detailed realization of these alternatives is left for future work.


In conclusion, the orange-triangle benchmark shows that the semi-annihilation process induced by the extra pseudo-scalar field can account for the observed DM abundance through thermal freeze-out and generate a neutrino signal compatible with the SK excess, while respecting self-interaction bounds. This motivates the analysis of the corresponding neutrino signal at the JUNO detector, presented next.
Although the best-fit points are not directly compatible with thermal production, we highlight that, $i)$ the orange triangle thermal benchmark lies within the $1\sigma$ confidence region, thus providing an excellent fit to the data, while $ii)$ alternative production mechanisms to re-populate the dark sector after freeze-out can be proposed, by invoking appropriate model-building strategies or modifications of standard cosmology, so that the best-fit points might remain a viable interpretation of the SK excess.

\subsubsection{Sensitivity of the JUNO detector}

The central JUNO detector consists of a liquid scintillator detector of 20 kton fiducial mass, designed to detect IBD with an excellent energy resolution of $3\%$, corresponding to $\sigma_E=0.03\,\text{MeV}\,\sqrt{E/\text{MeV}}$. The number of expected events per energy bin is given by Eq.~\eqref{eq:Nevents} where we used $E_{\nu}^{\rm min}=1.806\text{ MeV}$ \cite{JUNO:2015zny}, while the efficiency of the detector is taken from~\cite{JUNO:2022lpc}.

\begin{figure}[t]
  \centering
    \includegraphics[width=0.48\linewidth]{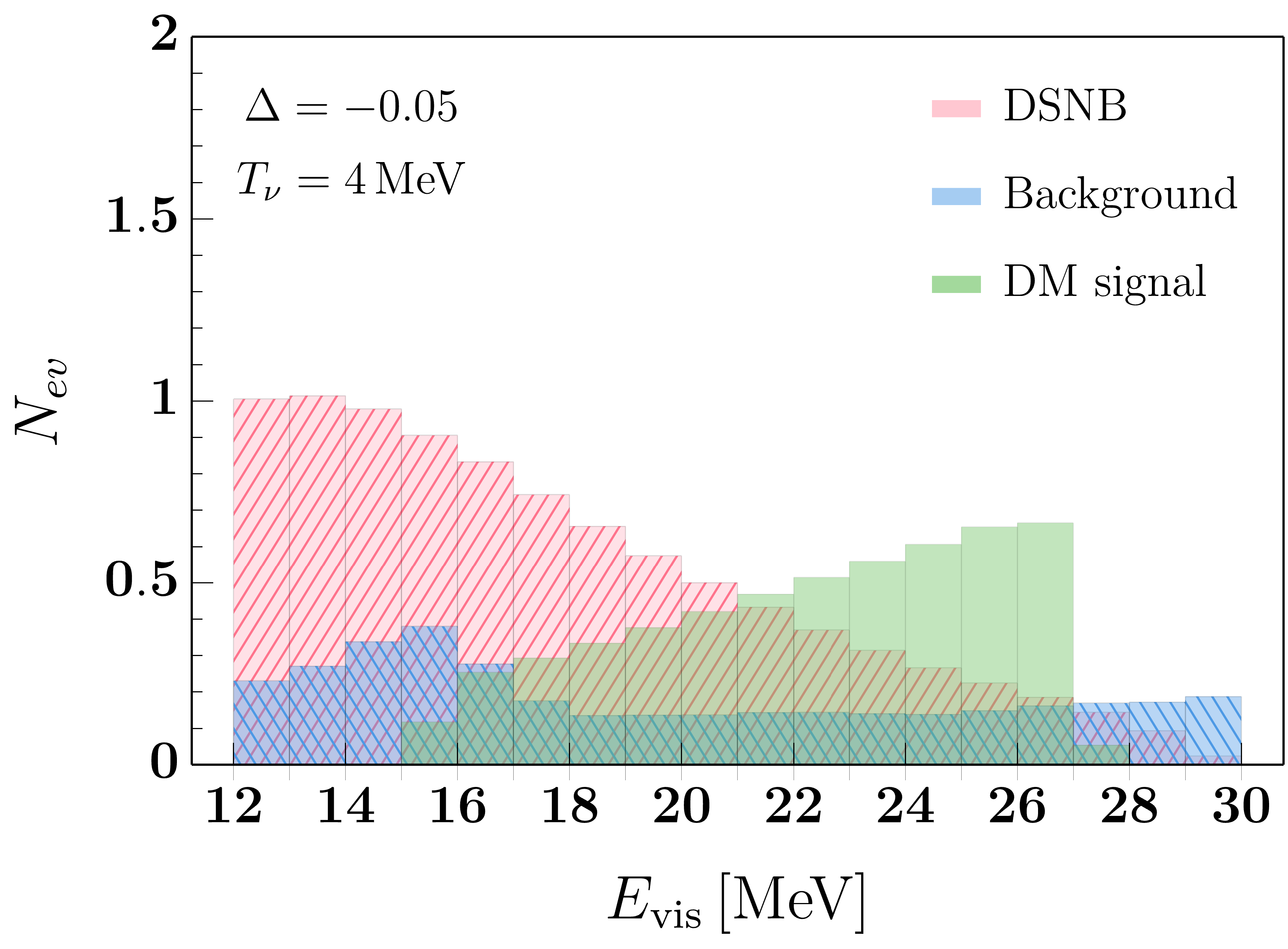}\includegraphics[width=0.48\linewidth]{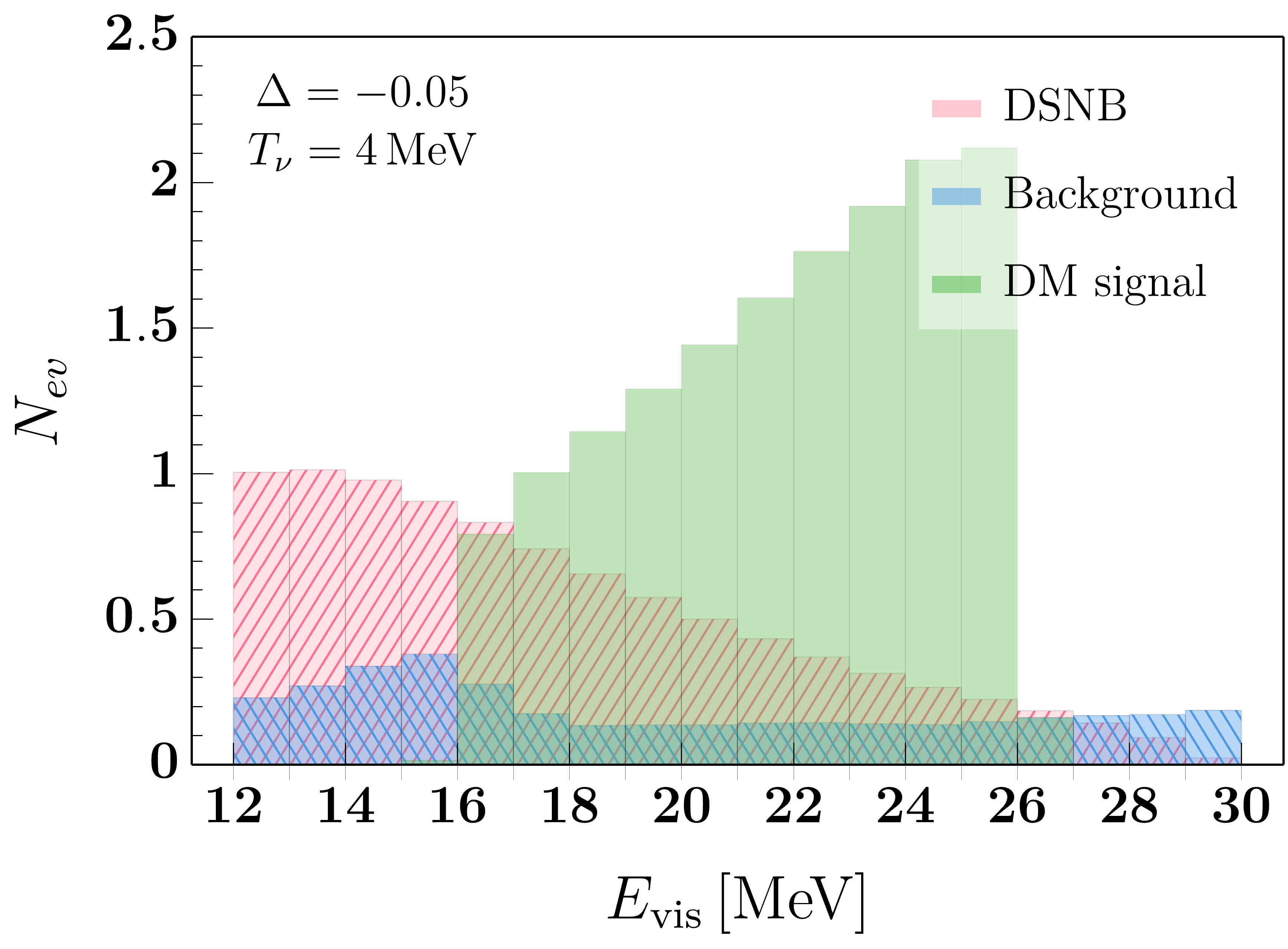}
    \caption{\textit{Left:} we show in green the predicted number of events per reconstructed-energy bin at JUNO for the representative benchmark marked by an orange triangle in Fig.~\ref{fig:SKmain}, compatible with thermal freeze-out production and self-interaction constraints. We also show the background in blue and the expected DSNB signal in red. See the main text for details. \textit{Right:} analogous results for the parameters corresponding to the orange star of Fig.~\ref{fig:SKmain}, for which thermal production is not viable.}
    \label{fig:JUNOmain}
\end{figure}

In Fig.~\ref{fig:JUNOmain} (left) we show in green the number of predicted events per energy bin for the benchmark point marked by the orange triangle in Fig.~\ref{fig:SKmain}, compatible with thermal production of DM. 
We consider the visible-energy interval $E_{\rm vis}\in[12,30]$ MeV and an exposure of 147 kton $\times$ yr.
The expected background is shown in blue and is taken from~\cite{JUNO:2022lpc}, while the expected DSNB contribution is shown in red.
To compute the latter, we digitized the differential DSNB flux from Fig. 4 of~\cite{Horiuchi:2008jz} (see also~\cite{Beacom:2010kk}). The authors show the DSNB flux for different choices of neutrino emission temperature.
For each choice, they display two curves 
representing the uncertainties resulting from the astrophysical inputs. We  take their mean at each energy as a representative prediction.
We adopted an optimistic value of 4 MeV for the neutrino emission temperature, which yields a relatively low DSNB background near the DM signal peak flux (in the $E_{\rm vis}\in[12,30]$ MeV range),   and thus provides an optimistic benchmark for DM searches. Analogous results for hotter spectra ($T=6$ MeV) are shown in Appendix~\ref{app:JUNOplots}. 

We estimate the expected signal significance using the Asimov approximation as
\begin{equation}
Z_A=\sqrt{2\sum_i\left[(N_i+B_i)\ln\left(1+\frac{N_i}{B_i}\right)-N_i\right]},
\end{equation}
where $N_i$ is expected signal  and $B_i$ the total background counts in the ($i$)-th energy bin, respectively, with the latter including the DSNB contribution. For an exposure of $147\,\text{kton}\times\text{yr}$,
the thermal benchmark marked by the orange triangle yields an expected significance of approximately $2\sigma$. 

The two best-fit points, marked by the two stars, yield much larger significances, around $5\sigma$, although these benchmarks require additional ingredients to populate the dark sector beyond standard thermal production, as discussed in the previous subsection. If a viable mechanism is provided, the resulting signal at the JUNO detector is greatly enhanced, as shown by the right panel of Fig.~\ref{fig:JUNOmain}, corresponding to the orange star of Fig.~\ref{fig:SKmain}.
These results illustrate JUNO's potential to provide an independent test of the DM interpretation of the SK excess.

\subsection{Constraints}\label{sec:constraints}

In this section we collect the existing constraints from perturbativity, cosmology, accelerator facilities and astrophysics.

\paragraph{Active-sterile mixings: constraints and prospects at future colliders.} The mixings among active and sterile neutrino species are constrained by several experiments and observations, including constraints from collider and beam-dump searches, Big Bang Nucleosynthesis (BBN), Cosmic Microwave Background (CMB) and neutrino facilities. See Ref.~\cite{Bolton:2019pcu} for a detailed review. For heavy sterile neutrinos of masses $M_{N}\gtrsim $ TeV, the strongest constraints are set by collider searches and electroweak precision data which require $\theta^2\lesssim 10^{-3}$. Furthermore, searches at future colliders such as FCC-hh, CLIC or ILC are expected to improve the constraints on the squared mixing angle up to one order of magnitude~\cite{Bolton:2019pcu}. 

\paragraph{Perturbativity.} The validity of ChPT requires $m_\pi\lesssim\Lambda\sim c_f f_\pi$. This is satisfied as long as $f_\pi\gtrsim 0.4\,m_\pi (2.7/c_f)$. Furthermore, the perturbative expansion in $\lambda_1$ coupling is valid as long as $y'_q\lesssim 0.9$.

\paragraph{BBN bound and $N_{\rm eff}$.} The pseudo-scalar mediator decays to active neutrinos with lifetime 
\begin{equation}
    \begin{split}
        \tau_a=1/\Gamma_a&\simeq2.2\times 10^{-12}\,\text{sec}\, \frac{1}{y_{Na}^2}\frac{50\,\text{MeV}}{m_a}\left(\frac{10^{-5}}{\theta^2}\right)^2\,,
    \end{split}
\end{equation}
well below 1 sec, thus avoiding bounds from Big Bang Nucleosynthesis (BBN). Notice that both the dark pions and the mediator are already deeply non-relativistic during neutrino decoupling. Their abundances are therefore strongly Boltzmann suppressed, and most of the dark-sector entropy is transferred while neutrinos are still thermally coupled to the electromagnetic plasma. Together with the short lifetime of $a$, which prevents a long-lived out-of-equilibrium population, this makes any modification of $N_{\rm eff}$ negligible.

\paragraph{CMB and LFV limits.} The interactions in Eq.~\eqref{eq:FundamentalLagrangian} induce $a\,\bar{\ell}(g_{a\ell\ell}^S+i\,g_{a\ell\ell}^P\gamma_5)\ell$ couplings of the pseudo-scalar $a$ to charged leptons at one loop (either through a loop of neutrinos attached to a $Z$ propagator or a triangle diagram with neutrinos and $W$ bosons running in the loop). For the inverse seesaw, the size of the couplings has been computed in~\cite{Herrero-Brocal:2023czw} (see also~\cite{Garcia-Cely:2017oco} for a discussion within the type-I seesaw mechanism),

\begin{equation}
    g_{a\ell\ell}\sim \frac{m_\ell m_\nu}{96\pi^2v^2}\sim g_{a\nu\nu}\left(\frac{m_\ell \mu}{96\pi^2v^2}\right)\,,
\end{equation}
where $g_{a\nu\nu}\sim\lambda_2\sim\theta^2$ denotes the coupling to active neutrinos.
These interactions induce DM semi-annihilations $\pi\pi\to \pi e^+e^-$. On the one hand, the electrons and positrons produced by DM annihilations can scatter off the photons of the Cosmic Microwave Background (CMB) and modify the evolution of recombination, leaving an imprint on the CMB anisotropies. On the other hand, X-ray telescopes can detect the photons produced via Inverse Compton scattering of background radiation off electrons originating from DM annihilations in halos. CMB measurements set the strongest bounds~\cite{Slatyer:2015jla}$, \langle{\sigma_{\pi e^+e^-}v_{\rm el}\rangle}\lesssim 5\times10^{-29}\,\text{cm}^3/\text{sec}$ for DM masses below $\sim$ 200MeV. In our scenario,

\begin{equation}
    \frac{\sigma_{\pi\,e^+e^-}}{\sigma_{\pi\,\bar{\nu}\nu}}\sim \left(\frac{m_e \mu}{96\pi^2v^2}\right)^2\sim\left(\frac{m_e m_\nu }{96\pi^2v^2\theta^2}\right)^2\sim10^{-31}\left(\frac{10^{-6}}{\theta^2}\right)^2\,,
\end{equation}
well beyond the current limit. Analogously, constraints from lepton flavor violating (LFV) processes mediated by $a$ are also safely evaded.

\paragraph{Supernovae and accelerator constraints.}

The $\lambda_2$ coupling in Eq.~\eqref{eq:pinulag} enables the production of pseudo-scalar mediators with energies $\sim 100$ MeV in a Supernova (SN) core through $\nu\nu\to a$ processes. The subsequent decay to (anti)neutrinos would produce a flux component with similar energies. The absence of $100$ MeV-range events in SN1987 signals imposes strong constraints. Focusing on masses in the  $m_a\sim 10-100$ MeV ballpark, the analysis in Ref.~\cite{Fiorillo:2022cdq} excludes values $2\times 10^{-11}(50\,\text{MeV}/m_a)\lesssim\lambda_2\lesssim2\times 10^{-9}(50\,\text{MeV}/m_a)$. For couplings larger than a few $\times\, 10^{-9}$, corresponding to $\theta^2\gtrsim 10^{-9}/y_N$, the pseudo-scalar $a$ decays inside the neutrinosphere, suppressing the high-energy neutrino signal underlying the SN1987A bound. Possible effects on supernova dynamics in this regime, particularly those associated with lepton-number violation, require a dedicated analysis.

Additional constraints on $m_a$ and $\lambda_2$ arise from accelerator searches for rare meson decays. Such constraints are summarized in~\cite{Foroughi-Abari:2026ofc}, which also discuss constraints from EW precision measurements. For values of $m_a$ in the $40-60$ MeV ballpark, the strongest constraints arise from pion decays (PIENU)~\cite{PIENU:2021clt} and kaon decays (NA62)~\cite{NA62:2021bji}, which require $\lambda_2\lesssim 5\times 10^{-3}$, weaker than the constraints derived from EW precision data on the active-sterile mixing angle.  

\paragraph{DM self-interactions.} Constraints from DM self-interactions are safely evaded as long as $f_\pi\gtrsim 28\,\text{MeV}\,(m_\pi/50\text{MeV})^{1/4}$. Since the DM relic abundance is now controlled by the Yukawa $y'_q$, those values are compatible with all the other constraints.

\paragraph{DM decays induced by Planck-suppressed operators.} 

Even though a global dark quark flavor symmetry prevents the DM particle from decaying, it is generically expected that quantum gravity effects break flavor universality, allowing the dark pions to decay through non-renormalizable Planck-suppressed operators. Assuming generic $\mathcal{O}(1)$ coefficients, following the analysis in~\cite{Garani:2021zrr}, the dark pions are safely stable on cosmological scales if $f_\pi\lesssim 5$ GeV\footnote{Planck-suppressed flavor-breaking operators may also induce mixing between some dark-pion states and the pseudo-scalar $a$, opening additional decay channels into neutrinos. Since the induced neutrino decay width scales as $\theta^4$, the constraints become substantially weaker for smaller active-sterile mixing. For sufficiently small allowed values of $\theta^2\lesssim 10^{-6}$, cosmological and indirect-detection bounds can therefore be satisfied for values of $f_\pi$ within the range considered in this work. The precise condition on the mixing angle and the Wilson coefficients of the non-renormalizable operators is generically model-dependent.}. Thus, combining constraints from perturbativity, DM decays and DM self-interactions, we found that DM masses of $\sim 50$ MeV are allowed for values of $28\,\text{MeV}\lesssim f_\pi\lesssim 5$ GeV, corresponding to $\mathcal{O}(100)\,\text{MeV}\lesssim \Lambda\lesssim \mathcal{O}(10)$ GeV. Notice that, in the absence of a primordial asymmetry, the cosmological abundance of dark baryons is negligible for such values of the parameters.

Finally, both the DM mass, through its dependence on $m_q$, and the active-sterile mixing angle $\theta^2$, depend on the same scalar vev $\langle\sigma\rangle$. This implies an extra condition on the parameter space, which can be expressed as $f_\pi\simeq 1.3\text{ GeV}(y_{N\sigma}/y_q)\,\left(m_\pi/50\text{MeV}\right)^2(\theta^2/10^{-6})$
The allowed window of $f_\pi$ corresponds to values of the mixing angle in the $2\times 10^{-8}\lesssim \theta^2\lesssim4\times 10^{-6}$ for $m_\pi\simeq 50$ MeV and $\mathcal{O}(1)$ Yukawa couplings $y_q\,,y_{N\sigma}$. Requiring the scalar mass $m_\sigma$ to be sufficiently large for it to be consistently integrated out of the low-energy theory restricts the allowed range of $f_\pi$. As a reference, scalar masses larger than 100 GeV (1 TeV) correspond to $f_\pi\gtrsim 300\text{ MeV}\,(1\text{ GeV})$, and, as a consequence, active-sterile mixing angles $\theta^2\gtrsim 2(8)\times 10^{-7}$. 

\vspace{0.3cm}

\section{A low-scale sterile variant}\label{sec:phenolight}

As anticipated at the end of Section~\ref{sec:simplemodel}, thermal DM production can also be reconciled with existing constraints without extending the particle content of Eq.~\eqref{eq:FundamentalLagrangian}. Although this scenario leads to heavier DM masses and is not intended to explain the SK excess, it can reproduce the observed relic abundance and provide complementary tests through sterile-neutrino searches at future accelerator experiments.

The key idea is to lower the sterile neutrino masses below the dark pion mass, $M_{N_{1,2}}\simeq M_D<m_\pi\lesssim\Lambda$, opening the annihilation channel $\pi\pi\to NN$. The corresponding interaction is controlled by the Majorana mass parameter $\mu$, rather than the light-neutrino masses, and therefore avoids the strong suppression of $\pi\pi\to\nu\nu$. This allows efficient DM depletion through thermal freeze-out, compatible with constraints from DM self-interactions.

To simplify the form of the interactions among the dark pions and the sterile neutrinos, we adopt a different choice of the mass matrices with respect to Eq.~\eqref{eq:benchmasses}, taking $\bm{\mu}=\mu\,\mathbf{1}_3$, $\bm{M}_D=M_D\,\mathbf{1}_3$ and $\bm{y}_\nu$ being the appropriate matrix to reproduce neutrino masses and mixings. The advantage of this choice, in particular the universal $\mu$ parameter, is to simplify the interactions among (active and sterile) neutrinos and dark pions. However, this is merely an illustrative choice and more generic structures would lead to similar results, with correction of $\mathcal{O}(1)$ for the relevant rates.  We assume the usual hierarchy of scales between $\mu,y_\nu v_H$ and $M_D$ so that the masses of active neutrinos are given in Eq.~\eqref{eq:numass}, while the steriles form a pseudo-Dirac pair (3 copies) with $M_{N_{1,2}}\sim M_D\pm \mu/2$.

At energies below $m_\sigma$, the interactions among dark pions and sterile neutrinos are described by 

    \begin{equation}
        \mathcal{L}_{\pi\pi NN}=\frac{\mu}{4N_ff_{\pi}^{2}}\pi_{i}^{2}\overline{N_{1}}N_{1}-\frac{\mu}{4N_ff_{\pi}^{2}}\pi_{i}^{2}\overline{N_{2}}N_{2}+\frac{\mu}{2N_ff_{\pi}^{2}}\pi_{i}^{2}\overline{N_{1}}i\gamma_{5}N_{2}\,.
    \end{equation}
    The annihilation channels $\pi\pi\to NN$ are now open, with total cross section (summing over all channels, $N_1N_1\,,N_2N_2\,,N_1N_2$)

   \begin{equation}
\ \sigma_{NN}\,v=\frac{3\mu^{2}}{4\pi N_f^2(N_f^2-1)f_{\pi}^{4}}\left(2-\frac{M_D^2}{m_\pi^2}\right)\sqrt{1-\frac{M_D^2}{m_\pi^2}}\,.
\end{equation}
In the limit $m_\pi\gg M_D$, the DM relic abundance is reproduced via thermal freeze-out for $ f_\pi\sim (M_{\rm Pl} T_{\rm eq}\mu^2)^{1/4}\,\sim {(M_{\rm Pl} T_{\rm eq}m_\nu^2/\theta^4)^{1/4}}$. More precisely, the numerical solution of the relevant Boltzmann equation for $N_f=2$ gives 

\begin{equation}\label{eq:fpilight}
  f_\pi\simeq 48\text{ GeV} \sqrt{\frac{10^{-10}}{\theta^2}}\,,  
\end{equation}
where, in this context,  $\theta^2\sim m_{\nu}/\mu\sim 0.05\text{ eV}/\mu$ is intended as an order of magnitude estimate for the physical mixings. With the large values of $f_\pi$ predicted by Eq.~\eqref{eq:fpilight}, constraints from DM self-interactions are safely evaded. 
Notice that the sterile neutrinos are kept in thermal equilibrium with the SM bath, at least until DM freeze-out, as long as  $y_\nu\gtrsim 10^{-8}$.

\begin{figure*}[t]
\centering
\includegraphics[width=0.5\textwidth]{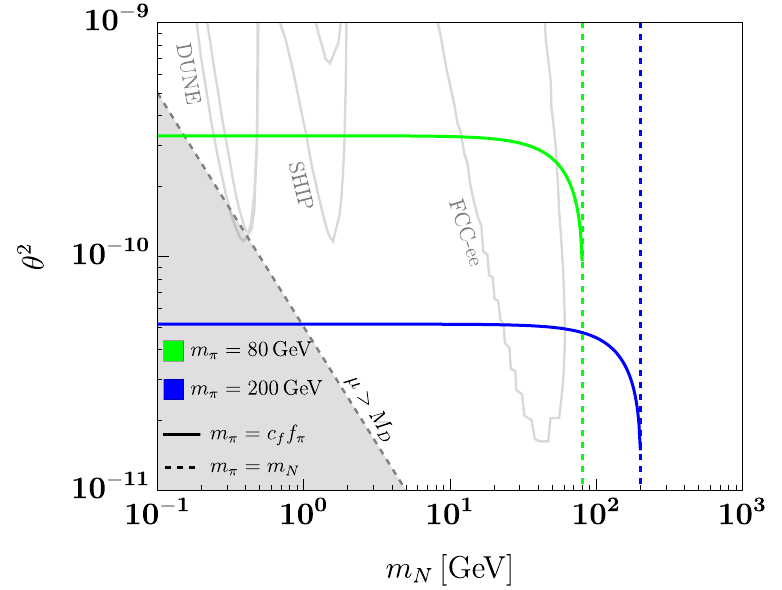} \,\,\,\,\,\,
\caption{
Values of active-sterile mixing angle as a function of the sterile mass for two different choices of DM mass and fixing $m_\pi=c_f f_\pi$. In the gray region, the inverse seesaw expansion breaks down. The dotted lines correspond to the kinematical endpoints $m_\pi=m_N$, where the annihilation processes become kinematically closed. The gray projections represent the sensitivities of proposed experiments and are taken from~\cite{Bolton:2019pcu}.
}\label{fig:lightsteriles}
 \end{figure*}

We show in Fig.~\ref{fig:lightsteriles} the predicted values for active-sterile mixing for $m_\pi=c_f f_\pi$, leaving the sterile mass as a free parameter. Within this assumption, for any value of the DM mass, the mixing is completely fixed by Eq.~\eqref{eq:fpilight}.
Displaced vertex searches for light steriles at the proposed FCC-ee collider  could potentially test mixings as small as $\theta^2\lesssim10^{-11}$~\cite{Antusch:2016vyf,Blondel:2022qqo,Abdullahi:2022jlv} for sterile masses in the $[20-60]$ GeV ballpark, while the SHIP proposal could test mixing $\theta^2\lesssim 10^{-10}$ for sterile masses of a few GeV~\cite{SHiP:2018xqw}.

Concerning constraints from indirect searches, Ref.~\cite{Batell:2017rol} and Ref.~\cite{Campos:2017odj} discuss the limits on DM particles annihilating into sterile neutrinos in dwarf spheroidal galaxies or the galactic center region. The steriles subsequently decay to charged leptons and SM pions, such as $N\to\nu\ell^+\ell^-$ or $N\to\pi^0\nu$, which produce gamma-ray photons that can be detected by FERMI-LAT, HESS telescopes, or antiprotons detected by AMS-02. Ref.~\cite{Batell:2017rol} excludes DM particles of mass below $m_\pi\lesssim80$ GeV (the precise number depends on the sterile mass, here we adopt the more conservative case). Ref.~\cite{Campos:2017odj} claims a stronger bound of $m_\pi\gtrsim 200$ GeV. We show our results for both $m_\pi=c_ff_\pi=\{80,200\}$ GeV in Fig.~\ref{fig:lightsteriles}.


Finally, for the heavy DM masses considered here, DM cosmological stability depends on the UV completion of the model. In particular, sufficiently long DM lifetimes are obtained if the leading Planck-suppressed decay operators have dimension six or higher, even for order-one Wilson coefficients, while dimension five operators could lead to dangerously short lifetimes if the corresponding Wilson coefficients are unsuppressed.


In conclusion, in this minimal setup, annihilations into sterile neutrinos can account for the observed DM relic abundance through thermal freeze-out, evading constraints from DM self-interactions. Searches for these states at future accelerator facilities, together with gamma-ray and neutrino searches at next-generation telescopes, could provide complementary tests of the scenario.

\section{Conclusions}\label{sec:conclusions}

We have studied the phenomenology of thermal dark-pion DM emerging from a new confining sector, which also produces small neutrino masses through the dynamical generation of a small lepton-number violating parameter from confinement dynamics and the presence of a scalar mediator. Interestingly, the same dynamics is responsible for both neutrino and DM mass generation.
DM annihilations into active neutrinos are suppressed by the active neutrino masses. Avoiding overproduction of DM requires a dark pion decay constant below the range allowed by DM self-interaction bounds, signaling a tension between cosmological and astrophysical data.

The introduction of an additional light pseudo-scalar particle, which could potentially be embedded together with the scalar mediator in a single complex field, opens a DM semi-annihilation channel to (anti)neutrinos. This can reconcile the thermal production of DM with existing constraints, selecting DM masses in the $\sim 10$ MeV range, and additionally providing a viable explanation for the $\sim 22$ MeV electron antineutrino excess reported by the SK collaboration. We perform a dedicated fit of the DM signal predicted by our model to SK data.
For a pseudo-scalar slightly lighter than the  DM, we find a benchmark that reproduces the observed relic abundance, satisfies self-interaction bounds, and yields a signal compatible with the SK excess at $1\sigma$. With an exposure of $147\,\mathrm{kton}\times\mathrm{yr}$, a dedicated analysis of searches for astrophysical neutrino fluxes at the JUNO detector could reach an expected significance of about $2\sigma$ for this thermal benchmark and $5\sigma$ for the best-fit points. An extra production mechanism to re-populate the dark sector after freeze-out could restore the observed DM abundance for the latter.

As an alternative way to reconcile the thermal production of DM with self-interaction constraints, without introducing extra degrees of freedom, we propose to lower the scale of the sterile neutrinos, so that the dark pions can annihilate as $\pi\pi\to NN$. While the selected DM masses lie above the GeV scale, and are not suitable to accommodate the SK excess, this scenario offers complementary tests through sterile-neutrino searches at future accelerators and through searches for DM annihilation products at gamma-ray and neutrino telescopes.

\section*{Acknowledgments}

We are deeply indebted to Salvador Rosauro-Alcaraz and Alessandro Granelli for exhaustive discussions on the SK dataset and the corresponding fitting procedure.
We would also like to thank Maximilian Berbig, Vasja Susic, Camilo Garcia-Cely and Federico Mescia for useful comments.  The work of GL is supported by the INFN Cabibbo Fellowship, call 2025. The work of JHG is supported by the Program of China Scholarship Council (Grant No. 202506020156). The work of CC is supported by PHENIICS Doctoral School (ED 576). JHG and CC acknowledge the Theoretical Phenomenology Visiting Institute at the Laboratori Nazionali di Frascati for hospitality during the early stage of the project.

\appendix

\section{Computation of $\pi\pi\to\pi\nu\nu$ cross section and neutrino spectrum}\label{app:23cross}

The Lagrangian in Eq.~\eqref{eq:pinulag} induces the semi-annihilation process $\pi_a \pi_b\to\pi_c \,\nu\nu\, (\pi_c\,\nubar\nubar)$, mediated by the exchange of an off-shell $a$.
We denote with $q$ the four-momentum of $a$ and neglect neutrino masses for simplicity.
Then, the amplitude of the process can be expressed as
\begin{equation}
|\mathcal{M}_{abc}|^2=\frac{36\times 8\lambda_1^2\lambda_2^2d_{abc}^2  q^2}{(q^2-m_a^2)^2}\,,
\end{equation}
where the factor 36 comes from $3!^2$ as we have 3 pion legs. To get the total cross section we need to sum over the final states and average over the initial ones. For $N_f=3$, we get 8 pions, so that $\sum_{a,b,c}d_{abc}^2={40}/{3}$, while the averaging gives a factor of $8^2=64$ in the denominator. Finally, we must multiply by the number of neutrino flavors ($3$) and divide by two to take into account that the final neutrinos are indistinguishable neutrinos.
Then, the cross-section times the relative velocity between the two initial pions is given by

\begin{equation}\label{eq:cross23}
    \sigma_{23}\,v_{}= \frac{45\lambda_1^2\lambda_2^2}{64\pi^3\,s^2}\int_0^{(\sqrt{s}-m_\pi)^2} dq^2 \frac{q^2\sqrt{\lambda_{\rm Kal}(s,m_\pi^2,q^2)}}{(q^2-m_a^2)^2}
\end{equation}
where $s\simeq4m_\pi^2$ is the usual Mandelstam variable, while $\lambda_{\rm Kal}(x,y,z)$ is the K\"{a}ll\'{e}n function. In the limit $q^2\ll m_a^2$, the integral can be performed analytically,
\begin{equation}
    \sigma_{23} v\simeq \frac{45\lambda_{1}^{2}\lambda^{2}_{2}\,m_{\pi}^{2}}{2048\pi^3 m_{a}^{4}}(57-40\ln4)= \frac{5 c_{f}^{6} {y'_{q}}^{2} y_{N}^{2} m_{\nu}^{2} m_{\pi}^{2}}{8192 \pi^3 m_{a}^{4} \mu^2}(57-40\ln4)\,. 
\end{equation}
On the other hand, close to the resonance threshold, $m_a\gtrsim m_\pi$, the expression in Eq.~\eqref{eq:cross23} is modified by replacing the $a$ propagator as $q^2-m_a^2\to (q^2-m_a^2)+im_a\Gamma_a$, where $\Gamma_a= \Gamma(a\to\nu\nu)+\Gamma(a\to\nubar\nubar)=3{\lambda_2^2m_a}/{4\pi}$ is the decay rate of the pseudo-scalar. In our computations and plots we take into account the Breit-Wigner corrected propagator and perform the integral in Eq.~\eqref{eq:cross23} numerically. Notice that for $m_a=m_\pi(1\pm \Delta)$ and close to resonance, $\Delta\ll1$, we get
\begin{equation}
    \frac{\Gamma_a^2m_a^2}{(q^2-m_a^2)^2}\overset{q^2\simeq m_\pi^2}{\simeq}\frac{\Gamma_a^2m_a^2}{4\Delta^2m_\pi^4}\simeq \frac{9\lambda_2^4}{64\pi^2\Delta^2}\simeq \frac{9y_N^4(m_\nu/\mu)^4}{1024\pi^2\Delta^2}\,.
\end{equation}
Since, in the interesting region of the parameter space, the maximal allowed value of the active-sterile mixing is $m_\nu/\mu\sim\theta^2\lesssim 10^{-3}$~\cite{Bolton:2019pcu}, and we consider $\Delta\gtrsim\mathcal{O}(0.01)$, the term proportional to $\Gamma_a^2$ is negligibly small.

\vspace{0.2cm}

The differential neutrino spectrum is obtained using the same amplitude
and phase-space factorization as above, but retaining the neutrino emission
angle $\theta^*$ instead of integrating over it, as we show below.

\paragraph{Kinematics of the neutrino boost.}
The off-shell pseudoscalar $a$ carries four-momentum $q=p_1+p_2-p_3$.  In the
center-of-mass (COM) frame of the initial pions, with $s=(p_1+p_2)^2$,
\begin{equation}
  E_q = \frac{s+q^2-m_\pi^2}{2\sqrt{s}}\,, \qquad
  |\mathbf{q}| = \frac{\sqrt{s}}{2}\,
    \lambda_{\text{Kal}}^{1/2}\!\!\left(1,\frac{m_\pi^2}{s},\frac{q^2}{s}\right).
\label{eq:qkin}
\end{equation}
In the rest frame of the emitted pseudo-scalar $a$, each neutrino has energy $E_\nu^*=\sqrt{q^2}/2$.
Boosting to the pion center-of-mass frame with $\gamma_q=E_q/\sqrt{q^2}$ and
$\beta_q=|\mathbf{q}|/E_q$, and denoting with $\theta^*$  the 
emission angle of the neutrino relative to the boost axis,
\begin{equation}
  E_\nu = \tfrac{1}{2}\bigl(E_q+|\mathbf{q}|\cos\theta^*\bigr).
\label{eq:Enu}
\end{equation}
Since $\cos\theta^*\in[-1,1]$, the neutrino energy is bounded by
\begin{equation}
  E_\nu^{\pm}(q^2)
  = \tfrac{1}{2}\bigl(E_q\pm|\mathbf{q}|\bigr)
  = \frac{1}{4\sqrt{s}}\!\left[
      \bigl(s+q^2-m_\pi^2\bigr)
      \pm s\,\lambda_{\text{Kal}}^{1/2}\!\!\left(1,\frac{m_\pi^2}{s},\frac{q^2}{s}\right)
    \right],
\label{eq:Enupm}
\end{equation}
so that $E_\nu^+-E_\nu^- = |\mathbf{q}|$.  Integrating
Eq.~\eqref{eq:Enu} over $\cos\theta^*\in[-1,1]$ (a factor of~2) recovers
Eq.~\eqref{eq:cross23}, providing a consistency check.

\paragraph{Neutrino energy spectrum.}
The total rate in Eq.~\eqref{eq:cross23} is obtained from the
doubly-differential rate
\begin{equation}
  \frac{\mathrm{d}^2(\sigma_{23}v_\mathrm{})}{\mathrm{d}q^2\,\mathrm{d}\!\cos\theta^*}
  = \frac{45\lambda_1^2\lambda_2^2}{128\pi^3 s}\,
    \frac{q^2\,\lambda_{\text{Kal}}^{1/2}\!\!\left(1,\dfrac{m_\pi^2}{s},\dfrac{q^2}{s}\right)}
         {(q^2-m_a^2)^2}
\label{eq:d2sig}
\end{equation}
by integrating $\cos\theta^*$ from $-1$ to $1$.  Instead, using
Eq.~\eqref{eq:Enu} to change variable to $E_\nu$ at fixed $q^2$,
\begin{equation}
  \left.\frac{\mathrm{d}\!\cos\theta^*}{\mathrm{d}E_\nu}\right|_{q^2}
  = \frac{2}{|\mathbf{q}|}\,
    \theta\!\left(E_\nu-E_\nu^-\right)\theta\!\left(E_\nu^+-E_\nu\right),
\end{equation}
and integrating over $q^2$, we obtain
\begin{equation}
  \frac{d(\sigma_{23}v_\mathrm{})}{dE_\nu}
  = \frac{45\lambda_1^2\lambda_2^2}{64\pi^3 s}
    \int_0^{(\sqrt{s}-m_\pi)^2}
    \frac{q^2\,\lambda_{\text{Kal}}^{1/2}\!\!\left(1,\dfrac{m_\pi^2}{s},\dfrac{q^2}{s}\right)}
         {(q^2-m_a^2)^2\,|\mathbf{q}|}
    \;\theta\!\left(E_\nu-E_\nu^-\right)\theta\!\left(E_\nu^+-E_\nu\right)
    \;\mathrm{d}q^2.
\label{eq:dsvdEnu}
\end{equation}

\section{Additional plots for JUNO event rate}\label{app:JUNOplots}

In this section we show the analogue of Fig.~\ref{fig:JUNOmain} for different choices of neutrino emission temperature corresponding to the DSNB spectra in Fig. 4 of~\cite{Horiuchi:2008jz}.  We also show in Table~\ref{tab:parameter_grid_combined} the statistical significance of the DM signal for each case, following the Asimov approach described in the main text.

\begin{table}
  \centering
  \begin{tabular}{lcc}
    \toprule
    \diagbox{$\Delta$}{$T_\nu$} & $4\text{ MeV}$ & $6\text{ MeV}$ \\
    \midrule
    $0.05\,\text{(best-fit)}$  & $4.9\sigma$ & $3.5\sigma$ \\
    \midrule
    $-0.05\,\text{(best-fit)}$ & $5.1\sigma$ & $3.7\sigma$ \\
    \midrule
    $-0.05\,\text{(triangle)}$   & $2.0\sigma$ & $1.4\sigma$ \\
    \bottomrule
  \end{tabular}
  \caption{Significances of the DM signal at JUNO for different benchmark points and values of $T_\nu$.}
  \label{tab:parameter_grid_combined}
\end{table}

\begin{figure}[H]
    \centering
    \includegraphics[width=0.45\linewidth]{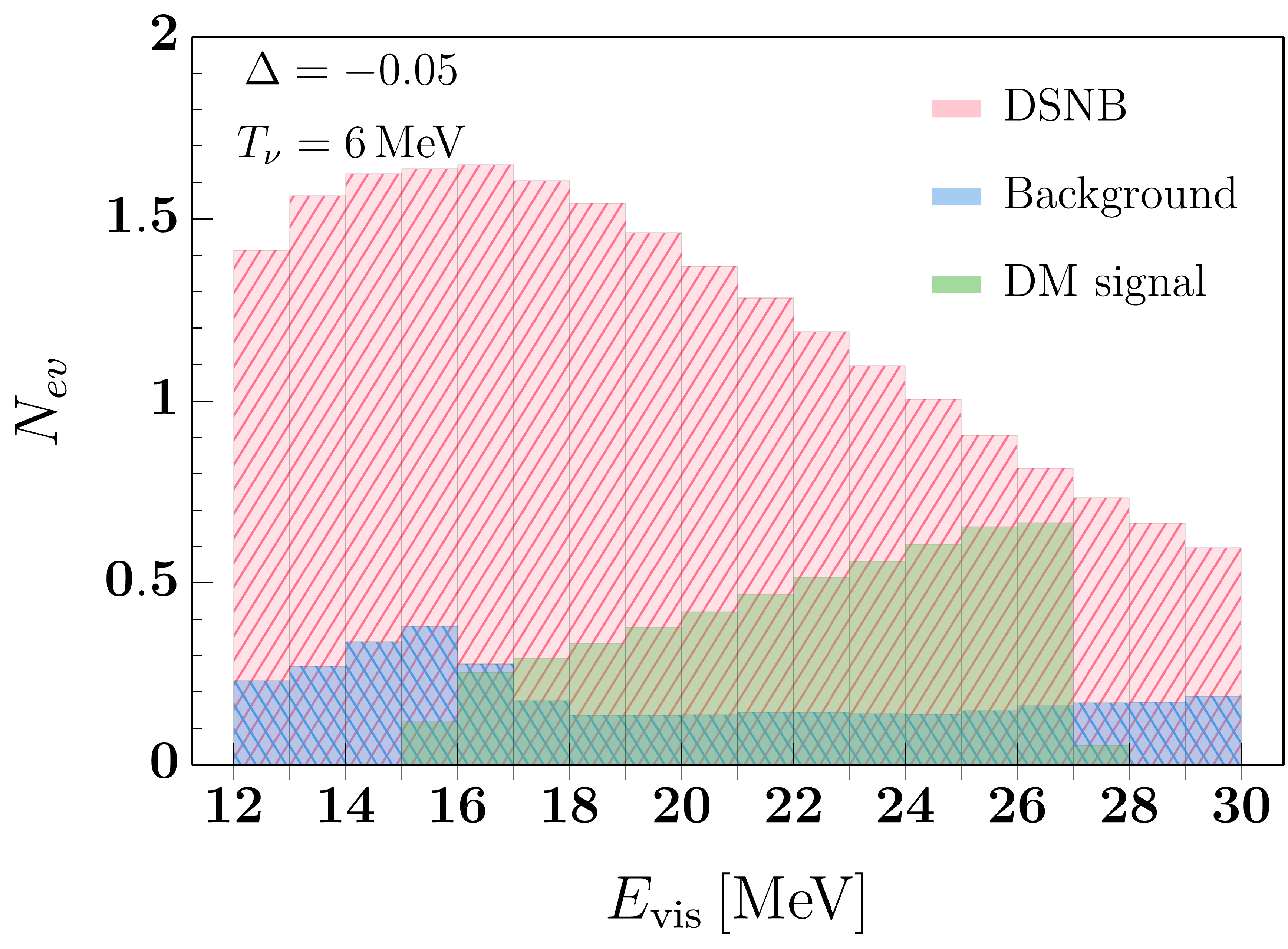}
    \includegraphics[width=0.45\linewidth]{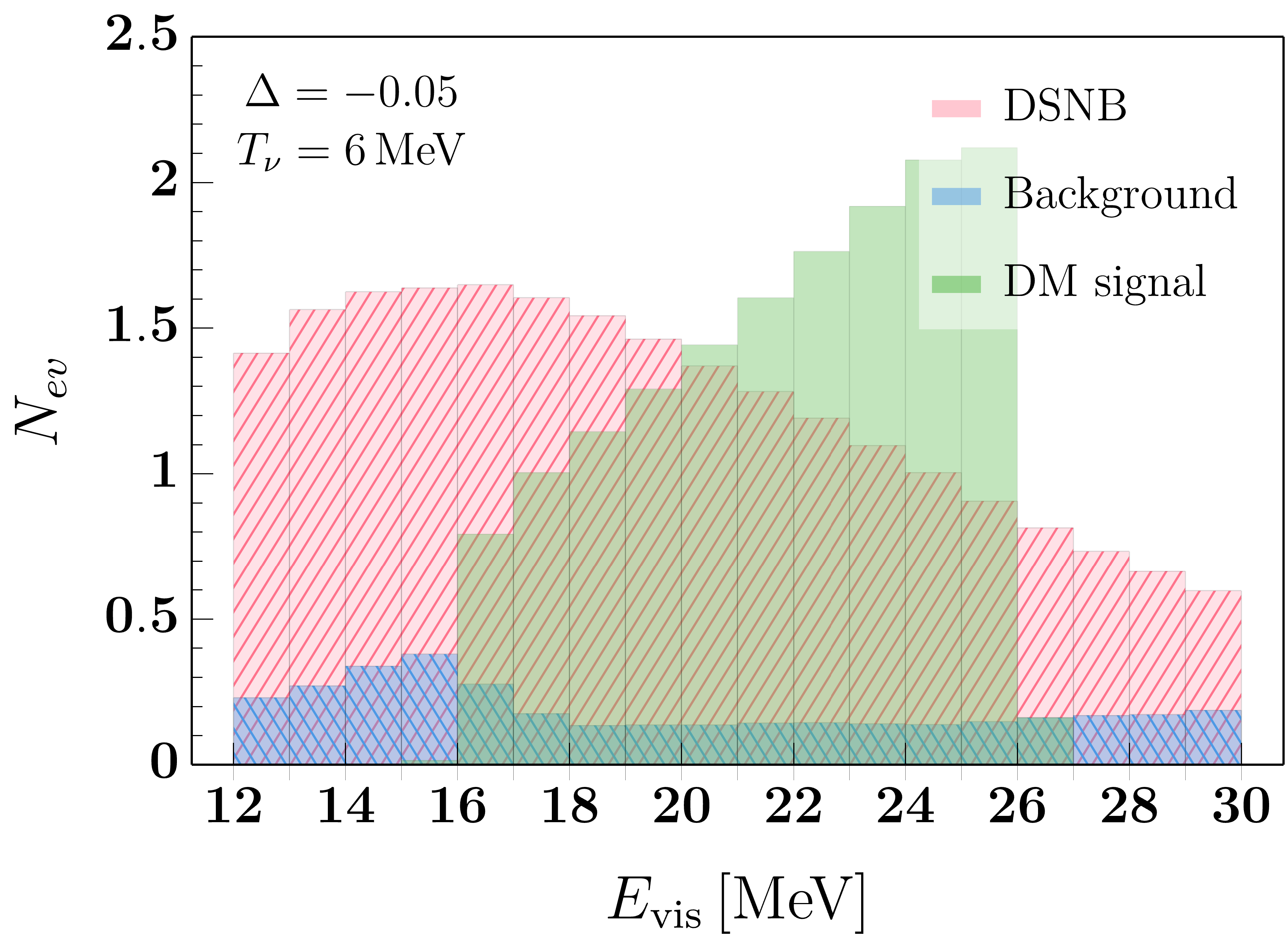}
     \\
     \centering  \includegraphics[width=0.45\linewidth]{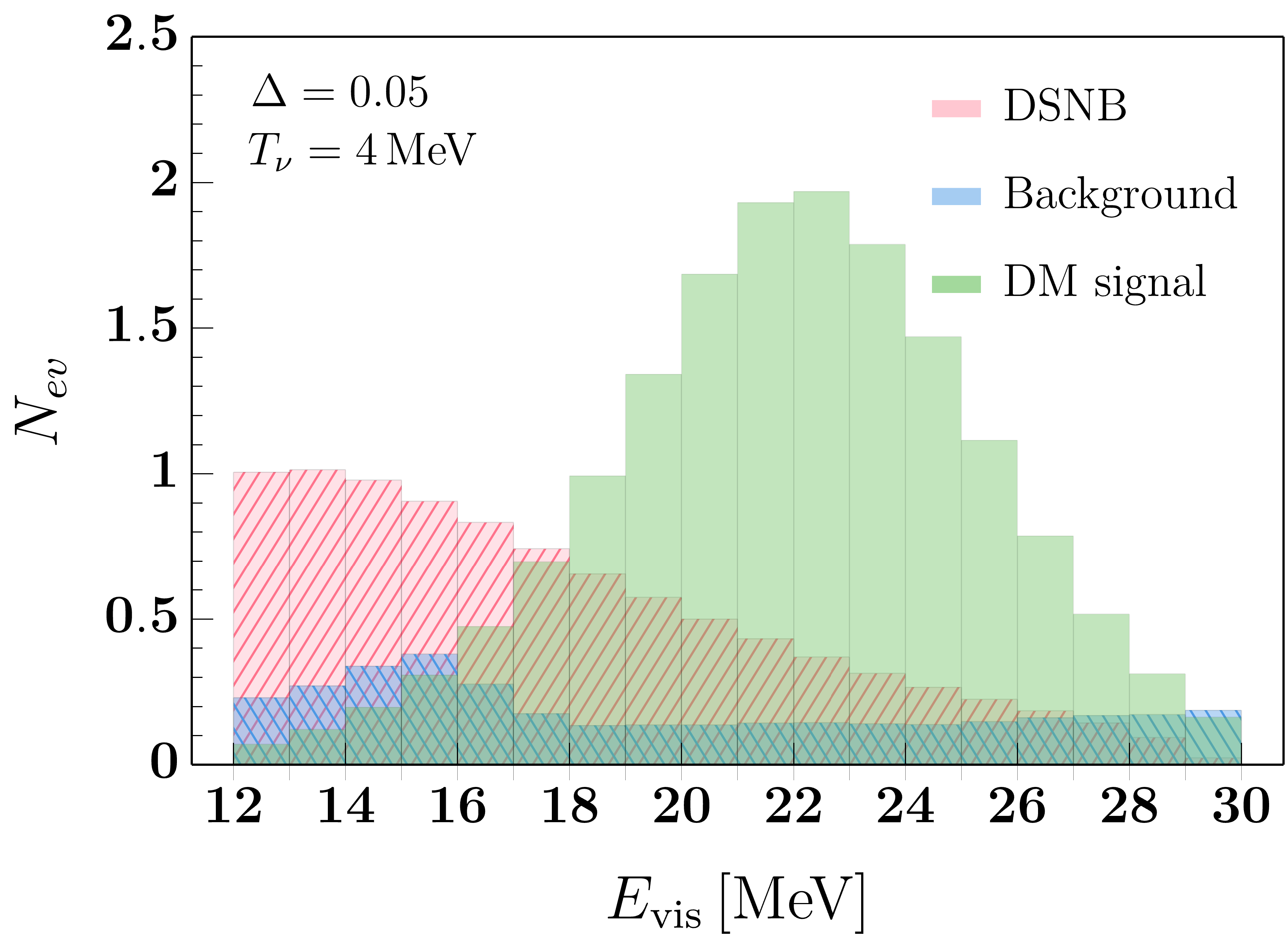}   
    \includegraphics[width=0.45\linewidth]{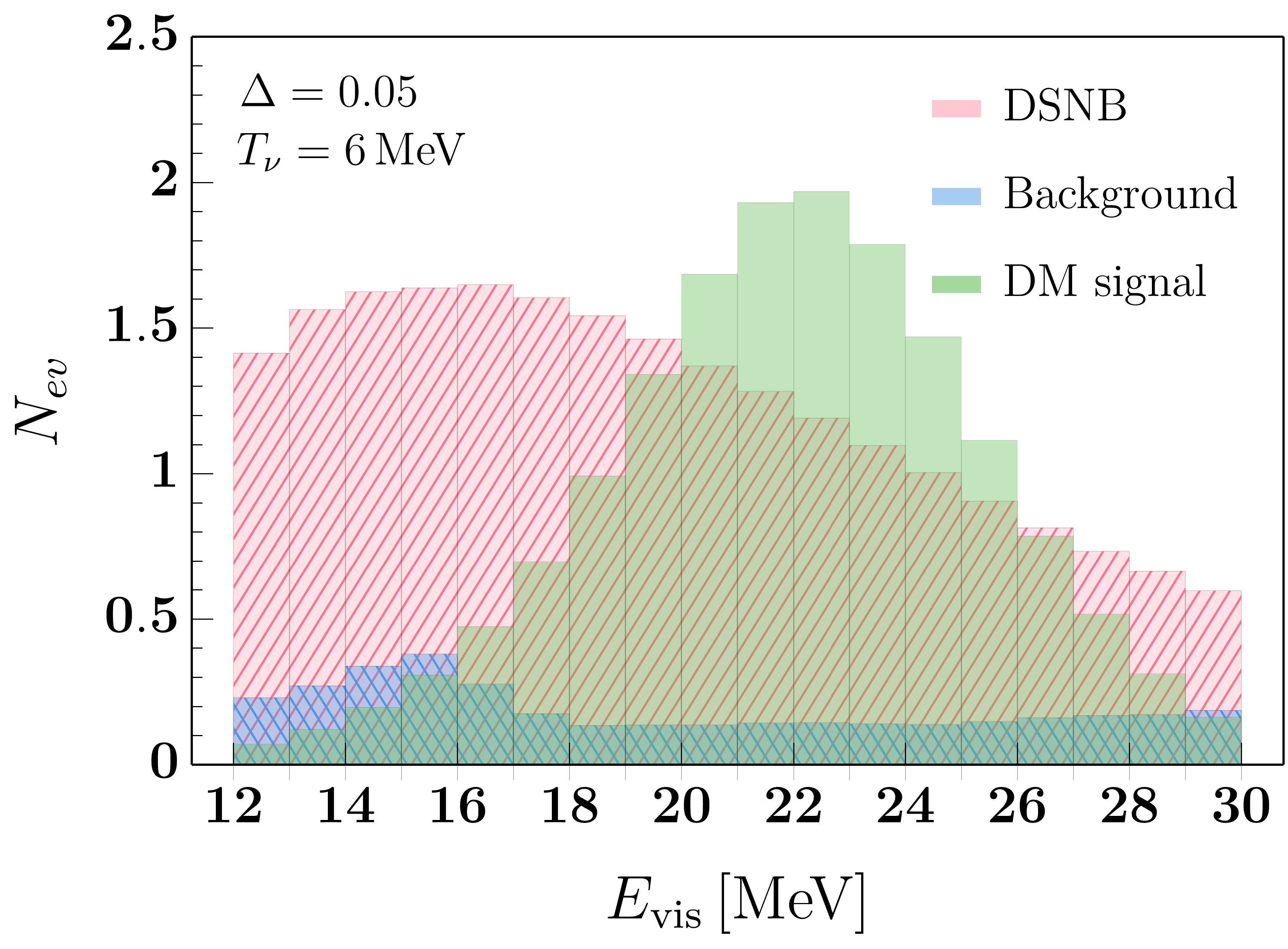}
    \caption{\textit{Upper panel:} same as Fig.~\ref{fig:JUNOmain} in the main text for a different choice of $T_\nu=6$ MeV. \textit{Lower panel:} Same plot for the best-fit point of Fig.~\ref{fig:SKmain} in the case $\Delta>0$, marked by a blue star.}
    \label{fig:placeholder7}
\end{figure}

\bibliographystyle{JHEP}
\bibliography{biblio}

\end{document}